\documentclass[a4paper]{amsart}

\usepackage[xindy]{glossaries}
\usepackage{amsfonts,mathtools,bm,amsthm, amssymb,bbm, mathrsfs}
\usepackage[OT1]{fontenc}
\usepackage[inline]{enumitem}
\usepackage{siunitx}
\usepackage{caption,subcaption}
\usepackage[hidelinks]{hyperref}
\usepackage[dvipsnames]{xcolor}
\usepackage{cleveref}
\usepackage{algorithm}
\usepackage{algpseudocode}

\theoremstyle{definition}

\newtheorem{example}{Example}

\theoremstyle{remark}

\DeclareMathOperator*{\argmax}{arg\,max}

\def\given{\,|\,}
\def\AND{\quad\text{and}\quad}
\def\tr{\intercal}
\def\op{o}

\title{Uncertainty modelling and computational aspects of data association}

\author[J. Houssineau]{Jeremie Houssineau}
\address{Department of Statistics of the University of Warwick, UK}
\email{jeremie.houssineau@warwick.ac.uk}
\author[J. Zeng]{Jiajie Zeng}
\address{Department of Statistics and Applied Probability, National University of Singapore}
\email{e0002186@nus.edu.sg}
\author[A. Jasra]{Ajay Jasra}
\address{Computer, Electrical and Mathematical Science and Engineering Division, King Abdullah University of Science and Technology, Thuwal, 23955-6900, KSA}
\email{ajay.jasra@kaust.edu.sa}

\begin{document}

\maketitle

\begin{abstract}
A novel solution to the smoothing problem for multi-object dynamical systems is proposed and evaluated. The systems of interest contain an unknown and varying number of dynamical objects that are partially observed under noisy and corrupted observations. An alternative representation of uncertainty is considered in order to account for the lack of information about the different aspects of this type of complex system. The corresponding statistical model can be formulated as a hierarchical model consisting of conditionally-independent hidden Markov models. This particular structure is leveraged to propose an efficient method in the context of Markov chain Monte Carlo (MCMC) by relying on an approximate solution to the corresponding filtering problem, in a similar fashion to particle MCMC. This approach is shown to outperform existing algorithms in a range of scenarios.
\end{abstract}


\section{Introduction}

We consider the problem of performing inference for multi-object dynamical systems under partial, corrupted and noisy observations. This class of problems, known as \emph{multi-target tracking} in the engineering literature \cite{Fortmann1980,Mahler2003,Vo2014}, arises in many applications, e.g.\ bio-imaging \cite{Chenouard2014}, robotics \cite{Mullane2011} and surveillance \cite{Benfold2011}, which can all benefit from principled inference solutions in different ways:
\begin{enumerate*}[label=\roman*)]
\item when the number of objects is too large to be treated by hand,
\item when the phenomena of interest take place on extended periods of time or, conversely, when an immediate response is needed,
\item when the data available about each object is scarce and
\item when it is difficult to tell one object from another.
\end{enumerate*}
One of the main difficulties with the considered type of system is that the number of objects is not known a priori and might vary in time due to a birth-death process. Also, objects are observed under multiple perturbations:
\begin{enumerate*}[label=\roman*)]
\item each object might or might not be detected,
\item if an object is detected then its state is only partially observed and the observation is subject to noise and
\item observations not related to any object, referred to as false alarms, are also received.
\end{enumerate*}
The main task when inferring the number of objects in a given system as well as their respective state is to solve the data association problem, that is, to estimate whether or not observations at different time steps originate from the same object. Each of the above-mentioned perturbations incurs a significant increase in the size of the set of all possible data associations, making it highly combinatorial. Due to this combinatorial nature, the task of estimating the current state of all objects based on all previous observations, referred to as multi-object filtering, is a difficult problem. It has been an active research topic for several decades and continues to be challenging in spite of the ever-increasing available computation resources \cite{Fortmann1980, Vo2014}. In this article, we aim to tackle the even harder problem of multi-object smoothing, that is, our objective is to keep evaluating the likelihood of data associations at previous times in light of newly received data. This is an important problem in practice since the elicitation of objects' trajectories and origins is fundamental for the evaluation of the objects' identities and of the associated situational awareness. Indeed, knowing the current state of each object is not sufficient in many situations and maintaining an up-to-date estimate of their past trajectories is often crucial. For instance, in defence applications, if an object labelled as ``ally'' crosses path with another object labelled as ``enemy'' then being able to tell one from the other at a later time can be more critical than having an accurate estimate of their state at that time.

In the context of filtering, one of the most natural ways of improving the trajectory estimates over the last few time steps is referred to as fixed-lag smoothing, where a sliding window made of a given number of time steps is updated based on the latest observations. The advantage with fixed-lag smoothing is that the computational cost can easily be tuned by selecting an adequate lag. However, since our objective is to elicit particular events that might have taken place at arbitrary time steps, we consider instead a ``batch'' alternative where a user-defined time-window of interest is fixed.

Defining a standard statistical model for representing multiple objects requires setting a number of probability distributions and parameters to characterise the different aspects of the problem, including highly uncertain phenomena such as false alarms. Such models also usually ignore the disparity between the different objects of interest in terms of behaviour and detection profile. In this article, we consider an alternative representation of uncertainty \cite{Houssineau2019_elements, Houssineau2018_parameter}, based on possibility theory \cite{Dubois2015}, that allows for acknowledging the lack of information about the different aspects of multi-object dynamical systems with the objective of increasing the robustness to misspecification of the derived solutions. The considered representation of uncertainty has links with imprecise probabilities \cite{Walley1991} and Dempster-Shafer theory \cite{Dempster1967,Shafer1976}. 

The use of MCMC to solve data association problems has been previously explored in \cite{Oh2009} as well as in \cite{Vu2014,Jiang2015,Jiang2018}. The approach considered in these articles is based on local proposals in the set of data association, with \cite{Vu2014,Jiang2015,Jiang2018} additionally considering the estimation of the object's trajectories. The objective in this article is to show that the set of data association can be explored effectively with global proposals without significantly affecting the probability of acceptance of each move. This result is achieved by leveraging the efficiency of an approximate multi-object filtering method. The use of MCMC in discrete spaces is discussed more generally in \cite{Zanella2019}. MCMC has also been used in conjunction with, or as a replacement of, sequential Monte Carlo in the context of filtering for multi-object systems, see e.g.\ \cite{Khan2005,Septier2009,Maroulas2012}; however this type of approach is less directly related to the method proposed in this article. 

Overall, the contributions of the articles are as follows:
\begin{enumerate*}[label=\roman*)]
\item a full multi-object model is defined in the context of possibility theory, building up on the components of single- and multi-object models of \cite{Ristic2019} and \cite{Houssineau2018_PHD};
\item a possibilistic analogue of a scalable solution to multi-object filtering \cite{Houssineau2018_TSP} is introduced;
\item the tools of possibility theory are used to define a suitable structure on the set of data associations;
\item a new efficient MCMC-based solution for the multi-object smoothing problem is introduced and its performance is demonstrated.
\end{enumerate*}

We introduce a new statistical model for representing multi-object systems in Section~\ref{sec:mainModel}. This is followed by the presentation of the proposed method for exploring the set of data association in Section~\ref{sec:dataAssociationMCMC}, before considering an extension of this approach in Section~\ref{sec:trackMCMC}. The performance of the proposed method is then assessed on simulated data in Section~\ref{sec:simulations}.

\section{Model}
\label{sec:mainModel}

We consider a fixed number $K$ of time steps and assume without loss of generality that time steps take integer values between $1$ and $K$. At each time step $k \in \{1,\dots,K\}$, a set of observations $Z_k$ is received, containing both object-originated observations and false alarms. Each observation in the set $Z_k$ is an element of an observation set $\mathsf{Z}$, which is assumed to be a subset of $\mathbb{R}^{d_{\mathsf{Z}}}$. In order to model that an object might not be detected, we introduce the notation $\phi$ for the empty observation, that is, an object for which detection has failed is associated with the empty observation $\phi$. We assume, as is standard, that an object cannot generate more than one observation at each time step. Therefore, denoting $\bar{Z}_k = Z_k \cup \{\phi\}$ the  set of observations at time $k$ augmented with the empty observation for any $k \in \{1,\dots,K\}$, any sequence of observations generated by an object through the $K$ time steps of the scenario can be seen as an element of
\begin{equation*}
\mathcal{O}_K = \bar{Z}_1 \times \dots \times \bar{Z}_K \setminus \{\phi\}^K
\end{equation*}
where the sequence of observation containing empty observations only is not considered. Elements of $\mathcal{O}_K$ are also referred to as \emph{observation paths} or simply as \emph{paths}. Data association can then be seen as the problem of determining the probability for all the paths in a given subset of $\mathcal{O}_K$ to be the true paths of objects in the system under consideration. Another standard assumption about multi-object systems is that each observation cannot originate from more than one object; as a consequence, not all subsets of $\mathcal{O}_K$ are considered feasible and we focus on the set $\mathcal{A}$ of subsets of $\mathcal{O}_K$ such that for all $A \in \mathcal{A}$, any two different observations paths $\op$ and $\op'$ in $A$ must verify that either $\op_k = \op'_k = \phi$ or $\op_k \neq \op'_k$ for all $k \in \{1,\dots,K\}$, where $\op_k$ denotes the $k$-th element of the sequence $\op$. Less formally, elements of $\mathcal{A}$ only contain paths that are different where they are not both equal to the empty observation. The set $\mathcal{A}$, in spite of being a strict subset of the power set of $\mathcal{O}_K$, has a large cardinality and evaluating the credibility of each of its elements by exhaustion can be difficult even when the number of observations at each time step is small. Assuming, for simplicity, that the number of observations at every time step is constant and equal to $m$, the number of elements in the power set of $\mathcal{O}_K$ is equal to $2^{m^K}-1$, which is prohibitively large even for toy problems. It is generally difficult to devise algorithms that perform inference on a large discrete space such as $\mathcal{A}$, yet, MCMC methods can help to address part of this challenge since they only require being able to evaluate the credibility of a given association $A \in \mathcal{A}$ proposed via some user-defined transition kernel.

In practice, we also need to estimate the interval of existence of each object. For this purpose, we introduce a set $\mathcal{T}$ which is similar to $\mathcal{A}$ except that each path $\op$ will be paired with a time of appearance $m \in \{1,\dots,K\}$ and the last time of existence $n \in \{m,\dots,K\}$. Formally, for all $T \in \mathcal{T}$, any $(\op,m,n)$ in $T$ must verify $\op_k = \phi$ for any $k \notin \{m,\dots,n\}$ and, for any $(\op',m',n')$ in $T$ different from $(\op,m,n)$, it must hold that either $\op_k = \op'_k = \phi$ or $\op_k \neq \op'_k$ for all $k \in \{1,\dots,K\}$, as for data associations. We denote by $\kappa$ the function extracting paths from tracks, that is $\kappa(t) = \op$ for any track $t$ with path $\op$.

\subsection{Uncertain variable and possibility function}

We consider a representation of uncertainty \cite{Houssineau2019_elements} which can be used as an alternative to subjective probabilities in a statistical model. The objective of this representation of uncertainty is to model information rather than randomness and therefore to address common issues with statistical modelling for complex systems and with the use of subjective probabilities. In the context of multi-object systems, some these issues are:
\begin{enumerate}[label=\arabic*)]
\item the associated models are inherently hierarchical which precludes the use of improper priors on the first level of this hierarchy; however, there is often no prior information on the location of appearing objects which means that uninformative priors are needed;
\item as with many complex systems, there is a large number of parameters which are not necessarily known in practice and learning these parameters is both challenging computationally as well as potentially useless if they are likely to change drastically from one time step to the other; this is for instance the case with the probability of detection;
\end{enumerate}
As will be shown in the next few sections, the proposed approach allows for addressing these issues while preserving most of the usual intuitive mechanisms in Bayesian inference.

We model a fixed but unknown quantity as a mapping $\bm{x}$ from a sample space $\Omega$ to a set $\mathsf{X}$, referred to as an \emph{uncertain variable}. The difference with a random variable is that $\Omega$ is not equipped with a probability distribution and, instead, there is a reference element in $\Omega$, denoted $\omega^*$, which correspond to the true value $x^* = \bm{x}(\omega^*)$ of the considered unknown quantity. The information about the true value of $\bm{x}$ is represented by a non-negative function $f_{\bm{x}}$ on $\mathsf{X}$ verifying $\sup_{x \in \mathsf{x}} f_{\bm{x}}(x) = 1$, referred to as a \emph{possibility function}. The scalar $f_{\bm{x}}(x) \in [0,1]$ corresponds to the credibility of the event $\bm{x} = x$ for any $x \in \mathsf{X}$ and the credibility of the event $\bm{x} \in A$ for any $A \subseteq \mathsf{X}$ is given by $\sup_{x \in A} f_{\bm{x}}(x)$. In particular, $f_{\bm{x}}$ is not a density and the integral is replaced by a supremum, which is consistent with the fact that the event $\bm{x} \in \mathsf{X}$ has credibility $1$ by construction. Possibility functions are not characterised by their corresponding uncertain variables and, instead, we say that the possibility function \emph{describes} the uncertain variable. If $\bm{y}$ is another uncertain variable in a set $\mathsf{Y}$ and if $\bm{x}$ and $\bm{y}$ are jointly described by the possibility function $f_{\bm{x},\bm{y}}$ then $\bm{y}$ is described by the marginal possibility function
$$
f_{\bm{y}}(y) = \sup_{x \in \mathsf{X}} f_{\bm{x},\bm{y}}(x,y), \qquad y \in \mathsf{Y},
$$
and the possibility function describing $\bm{x}$ given that $\bm{y} = y$ is
$$
f_{\bm{x}}(x \given y) = \dfrac{f_{\bm{x},\bm{y}}(x,y)}{f_{\bm{y}}(y)} = \dfrac{f_{\bm{y}}(y \given x)f_{\bm{x}}(x)}{\sup_{x' \in \mathsf{X}} f_{\bm{y}}(y \given x')f_{\bm{x}}(x')}, \qquad x \in \mathsf{X},
$$
which is the analogue of Bayes' theorem for possibility functions \cite{deBaets1999}. In this context, we will refer to $f_{\bm{x}}$ and $f_{\bm{x}}(\cdot \given y)$ as the prior and posterior possibility functions respectively and $f_{\bm{y}}(y \given \cdot)$ will be called the likelihood function; similarly, $f_{\bm{y}}(y)$ will be referred to as the marginal likelihood. If it holds that $f_{\bm{x},\bm{y}}(x,y) = f_{\bm{x}}(x)f_{\bm{y}}(y)$ for all $(x,y) \in \mathsf{X}\times\mathsf{Y}$ then $\bm{x}$ and $\bm{y}$ are said to be independently described. This form of independence only implies that the information about $\bm{x}$ is not related to the information we hold about $\bm{y}$.

The expected value and variance can be defined for possibility functions via the corresponding law of large numbers and central limit theorem \cite{Houssineau2019_elements} as
\begin{align*}
\mathbb{E}^*(\bm{x}) & = \argmax_{x \in \mathsf{X}} f_{\bm{x}}(x) \\
\mathbb{V}^*(\bm{x}) & = \big(- \Delta f_{\bm{x}}\big(\mathbb{E}^*(\bm{x})\big) \big)^{-1} = \mathbb{E}^*\big(-\Delta \log f_{\bm{x}}(\bm{x}) \big)^{-1},
\end{align*}
where $\Delta f_{\bm{x}}$ is the Laplacian of $f_{\bm{x}}$, with the variance being infinite when $\mathbb{E}^*(\bm{x})$ is not a singleton and undefined when $f_{\bm{x}}$ is not twice differentiable at $\mathbb{E}^*(\bm{x})$. The variance can be seen as being the inverse of an analogue of the Fisher information. Another useful notion of expected value, which is the direct analogue of the standard expected value, can be defined for any real-valued function $\varphi$ on $\mathsf{X}$ as
$$
\bar{\mathbb{E}}(\varphi(\bm{x})) = \sup_{x \in \mathsf{X}} \varphi(x) f_{\bm{x}}(x).
$$
The scalar $\bar{\mathbb{E}}(\varphi(\bm{x}))$ can be interpreted as the maximum expected value of $\varphi(\bm{x})$.

Many concepts and results holding for probability distributions can be used for possibility functions. For instance, if $\mathsf{Y} = \mathsf{X} = \mathbb{R}$ and if the likelihood function is a normal possibility function, i.e.\
$$
f_{\bm{y}}(y \given x) = \exp\Big( -\dfrac{1}{2\sigma^2}(y - a x)^2 \Big) \doteq \overline{\mathrm{N}}(y; a x, \sigma^2)
$$
for some $a \in \mathbb{R}$ and some $\sigma > 0$, then one can show that the posterior is also a normal possibility function if the prior $f_{\bm{x}}$ is normal. In other words, the concept of conjugate priors makes sense. This result can be extended to the multivariate case and it has been shown in \cite{Houssineau2018_UQ} that the posterior expected value and variance of the Kalman filter can be recovered with possibility functions.

If the objective is to find the (subjective) probability $p(B)$ of some event $\bm{x} \in B$ for some measurable subset $B$ of $\mathsf{X}$, then the credibility $\sup_{x \in B} f_{\bm{x}}(x)$ can be seen as an upper bound for this probability and we find that
\begin{equation}
\label{eq:subjective_probability}
1 - \sup_{x \in B^{\mathrm{c}}} f_{\bm{x}}(x) \leq p(B) \leq \sup_{x \in B} f_{\bm{x}}(x),
\end{equation}
where $B^{\mathrm{c}} = \mathsf{X} \setminus B$ is the complement of $B$ in $\mathsf{X}$. This interpretation implies that the possibility function $\bm{1}$, which is equal to $1$ everywhere on $\mathsf{X}$, is the least informative. This uninformative possibility function is well defined even when $\mathsf{X}$ is unbounded. It is also possible to interface uncertain variables and random variables in order to introduce more sophisticated representations of uncertainty involving both lack of information and randomness \cite{Houssineau2018_parameter}. However, we will argue that all the elements of the introduced statistical model can be seen as subjective so that only possibility functions will be used.

\subsection{Multi-object model}
\label{sec:model}

We first introduce the assumptions and notations for modelling the way objects appear, behave and disappear in Section~\ref{sec:dynamics} before moving on to the considered sensor modelling in Section~\ref{sec:observation}. Most of the assumptions are standard in the field of multi-object estimation.

\subsubsection{Object and population dynamics}
\label{sec:dynamics}

We consider the case where there is no information about some or all of the components of the state of appearing objects. Typically, there might be no prior information about the position of objects whereas assumptions can be made about the velocity components. Denoting $m \in \{1,\dots,K\}$ the time step at which a given object has appeared, the state at this time step is represented by an uncertain variable $\bm{x}_m$ in a space $\mathsf{X} \subseteq \mathbb{R}^{d_{\mathsf{X}}}$ described by a possibility function $f_0$. With probabilistic modelling, improper priors might be required in order to model the absence of information about appearing objects; however, the hierarchical nature of multi-object estimation implies that improper priors cannot be used without adding heuristics at the level of data association \cite{Houssineau2010, Ristic2012}.

We consider that there is a non-negligible heterogeneity between the dynamics of the different objects and that the characteristics of the objects' motion is not necessarily well known. As a consequence, we model the trajectory of an object as a sequence of uncertain variables $\{\bm{x}_k\}_{k=m+1}^n$ on $\mathsf{X}$ such that, for any $k \in \{m+1,\dots,n\}$, $\bm{x}_k$ is described by a possibility function $f_{\bm{x}_k}(\cdot \given \bm{x}_m, \dots, \bm{x}_{k-1})$ satisfying
$$
f_{\bm{x}_k}(x_k \given \bm{x}_m, \dots, \bm{x}_{k-1}) = g_k(x_k \given \bm{x}_{k-1}), \qquad x_k \in \mathsf{X},
$$
for some possibility function $g_k(\cdot\given\bm{x}_{k-1})$ on $\mathsf{X}$. This is an analogue of the Markov property for uncertain variables.

We take into account the fact that objects might completely disappear from the scene before the last time step, in which case we say that the object has ``not survived''. This could be seen as a convenient way of dealing with objects that are no longer detectable by the sensor(s). Object survival is not usually a random event so that we model it as an uncertain variable. The respective credibilities for an object with state $x \in \mathsf{X}$ to survive or not survive to the next time step are denoted $\alpha_{\mathrm{s}}(x)$ and $\alpha_{\mathrm{ns}}(x)$. These credibilities must verify $\max\{\alpha_{\mathrm{s}}(x), \alpha_{\mathrm{ns}}(x)\} = 1$ for any $x \in \mathsf{X}$. We consider the case where $\alpha_{\mathrm{s}} = \bm{1}$ since we want to model that objects are unlikely to disappear right after appearing, for which we need to set $\alpha_{\mathrm{ns}}(x) \ll 1$ for any $x \in \mathsf{X}$. The subjective probability of survival for an object with state $x \in \mathsf{X}$ is therefore restricted to the interval $[1-\alpha_{\mathrm{ns}}(x), 1]$.

Given the introduced model and notations, the joint credibility of a trajectory $x_{m:n} \in \mathsf{X}^{n-m+1}$ and of the corresponding last time of existence $n \in \{1,\dots,K\}$ for an object that is known to appear at time step $m$ can be characterised by the possibility function
$$
g(x_{m:n},n \given m) = f_0(x_m) \alpha_{\mathrm{ns}}(x_n)^{\mathbbm{1}(n < K)} \prod_{k = m+1}^n g_k(x_k \given x_{k-1}),
$$
where $\mathbbm{1}(e)$ equals $1$ if $e$ is true and $0$ otherwise. 

There are several possible models for the number of appearing objects per time step. The simplest is to assume that the credibility for an object to appear at time $k \in \{1,\dots,K\}$ is $\alpha_{k,+}$ and that this aspect can be independently described for all objects. The credibility for $M$ objects to appear at time step $k$ is then equal to $\alpha_{k,+}^M$. Additional information might however be available about appearing objects, such as a maximum number $M_{k,+}$ at time step $k$, in which case we would have a credibility of $\mathbbm{1}(M \leq M_{k,+})\alpha_{k,+}^M$. The associated possibility function on the set $\mathbb{N}_0$ of non-negative integers is denoted $f_{k,+}$ in any case.

\subsubsection{Observation}
\label{sec:observation}

Most sensors acquire information about the objects of interest by measuring some signal over an array of resolution cells. This is the case for cameras, where these resolution cells are pixels, but also for most radars and sonars \cite{Skolnik1990}. Considering for instance the case of a radar measuring range and azimuth, the internal processing of the radar image yields a set of resolution cells where the strength of the signal suggests the presence of an object in the corresponding directions and at the specified distances. In addition, objects are often extended and the signal can originate from different edges and/or surfaces depending on their (unknown) orientations. As a consequence, we model the observation process via uncertain variables and consider the following form for the likelihood function:
\begin{equation*}
\ell_k(z \given x) = \exp\Big( -\dfrac{1}{2}(z - h_k(x))^{\tr}R_k^{-1}(z - h_k(x)) \Big) \doteq \overline{\mathrm{N}}(z; h_k(x), R_k),\qquad  z \in \mathsf{Z},
\end{equation*}
where $R_k$ is a $d_{\mathsf{Z}}\times d_{\mathsf{Z}}$ symmetric positive-definite matrix related to the size and shape of the resolution cells (assumed constant in $\mathsf{Z}$). The difference between this normal possibility function and the corresponding normal probability distribution would not matter in a standard single-object tracking scenario since normalising constants would simplify in Bayes' theorem; however, in multi-object tracking, these constants are important since they appear in the assessment of data associations. The credibility for an object with state $x \in \mathsf{X}$ to be detected is denoted $\alpha_{\mathrm{d}}(x)$ and, similarly, the credibility of a detection failure is denoted $\alpha_{\mathrm{nd}}(x)$. Since it must hold that $\max\{\alpha_{\mathrm{d}}(x), \alpha_{\mathrm{nd}}(x)\} = 1$ for any $x \in \mathsf{X}$, we will assume that $\alpha_{\mathrm{d}} = \bm{1}$ so that it is unlikely for an object to remain undetected. Given a trajectory $x_{m:n}$ of an object appearing at time step $m$ and disappearing after time step $n$, it follows that the likelihood function for a path $\op \in \mathcal{O}_K$ is
\begin{equation*}
\ell(\op \given x_{m:n}, m, n) = \prod_{k = m}^n \alpha_{\mathrm{nd}}(x_k)^{\mathbbm{1}(\op_k = \phi)} \ell_k(\op_k \given x_k)^{\mathbbm{1}(\op_k \neq \phi)}.
\end{equation*}

The credibility for an observation $z \in \mathsf{Z}$ at time $k \in \{1,\dots,K\}$ to be a false alarm is denoted $\alpha_{k,\mathrm{fa}}(z)$, which will be assumed to be strictly lesser than $1$; otherwise, if it were possible for all observation to be false alarms then this would be the posterior expected data association in general. The credibility for a given finite subset $Z$ of observations in $\mathsf{Z}$ to be false alarms is then
\begin{equation*}
f_{k,\mathrm{fa}}(Z) = \prod_{z \in Z} \alpha_{k,\mathrm{fa}}(z).
\end{equation*}
As a possibility function on sets, $f_{k,\mathrm{fa}}$ must verify that $\sup_{Z \subseteq \mathsf{Z}} f_{k,\mathrm{fa}}(Z) = 1$.

\subsection{Target possibility function}
\label{sec:smoothing}

We now introduce the posterior possibility function on the set $\mathcal{T}$ describing the unknown set of tracks, based on the model detailed in Section~\ref{sec:model}. For this purpose, we consider a track $t = (o, m, n)$ and start by defining the credibility $\pi(\op,n \given m)$ of the pair $(\op,n)$ given the time of appearance $m \in \{1,\dots,K\}$ as
\begin{equation*}
\pi(\op,n \given m) = \sup \big\{ \ell(\op \given x_{n:m}, m, n) g(x_{n:m}, n \given m) : x_{n:m} \in \mathsf{X}^{n-m+1}  \big\}.
\end{equation*}
Other aspects such as false alarms and initial observations must be considered jointly. We denote $f_{\mathrm{fa}}$ the possibility function defined on $\mathcal{A}$ as
$$
f_{\mathrm{fa}}(A) = \prod_{k = 1}^K f_{k,\mathrm{fa}}(Z_{k,\mathrm{fa}}(A)),
$$
for any $A \in \mathcal{A}$, where $Z_{k,\mathrm{fa}}(A) = \{ z \in Z_k : \forall \op \in A, z \neq \op_k \}$ is the set of false alarms induced by $A$ at time step $k$. We also introduce $f_+$ as the possibility function on $\mathcal{T}$ defined as
$$
f_+(T) = \prod_{k = 1}^K f_{k,+}(M_k(T))
$$
where $M_k(T) = \sharp\{ (\op,m,n) \in T : m = k \}$ is the number of objects appearing at time step $k \in \{1,\dots,K\}$. The functions $f_{\mathrm{fa}}$ and $f_+$, defined respectively on $\mathcal{A}$ and $\mathcal{T}$, are not possibility functions; instead, they are simply the joint credibility for observations that are not in a given element of $\mathcal{A}$ to be false alarms and for tracks that are in a given element of $\mathcal{T}$ to have appeared at the indicated time steps. The target possibility function, i.e.\ the posterior possibility function $\Pi$ on $\mathcal{T}$ describing the unknown set of tracks, is then expressed for any $T \in \mathcal{T}$ as
\begin{equation}
\label{eq:smoothingTrack}
\Pi(T) \propto f_{\mathrm{fa}}(\kappa(T)) f_+(T) \prod_{(\op,m,n) \in T} \pi(\op,n \given m ),
\end{equation}
and is such that $\max_{T \in \mathcal{T}} \Pi(T) = 1$. The marginal likelihood for the set of paths $A = \kappa(T)$ is then defined as
\begin{equation}
\label{eq:smoothingPath}
\hat{\Pi}(A) = \max\{ \Pi(T) : T \in \mathcal{T}, \kappa(T) = A \}.
\end{equation}

\section{MCMC for data association}
\label{sec:dataAssociationMCMC}

\subsection{Computational aspects of possibility theory}

Approximation methods for possibility functions must be devised in order to solve the corresponding inference problems in general. Grid-based methods have the same shortcomings as in the probabilistic case since it is often difficult to anticipate where the posterior possibility function will take non-negligible values. Although one cannot sample directly from a given possibility function $f_{\bm{x}}$, the latter can be used within MCMC together with a proposal (probability) distribution. In this case, there is no requirement of targeting a given probability distribution and there is no concern regarding the independence between samples. One of the consequences is that low-discrepancy sequences can be used instead of pseudorandom numbers. The generated chain, say $\{X_n\}_{n \geq 1}$, will simply be used to approximate the expected value
$$
\bar{\mathbb{E}}(\varphi(\bm{x})) \approx \max_{n \geq 1} \varphi(X_n) f_{\bm{x}}(X_n),
$$
for any real-valued function $\varphi$ on $\mathsf{X}$. As opposed to the standard Monte Carlo approximation, the possibility function $f_{\bm{x}}$ appears explicitly in the expression of $\bar{\mathbb{E}}(\varphi(\bm{x}))$ since the density of samples in a given area conveys no information about $f_{\bm{x}}$; instead, the chain $\{X_n\}_{n \geq 1}$ simply provides support points for the approximation of $f_{\bm{x}}$ as a function. If only the expected value $\mathbb{E}^*(\bm{x})$ of $\bm{x}$ is of interest, then the possibility function $f_{\bm{x}}^{\gamma}$ for some $\gamma > 1$ can be used instead. The considered power can also be increased during the execution of the MCMC, leading to a simulated annealing. Conversely, if one is interested in identifying the subset of $\mathsf{X}$ containing at least $100(1-\alpha)\%$ of the subjective probability mass defined in \eqref{eq:subjective_probability}, then areas where $f_{\bm{x}}$ has value $\alpha$ must also be explored, hence justifying the use of a power $\gamma$ strictly lesser than $1$.

When using the possibility function $f^{\gamma}_{\bm{x}}$ in a MCMC algorithm, it is the probability distribution on $\mathsf{X}$ defined as the renormalised version of $f^{\gamma}_{\bm{x}}$ that is targeted (assuming $f^{\gamma}_{\bm{x}}$ is integrable). This is not however the only possible approach. Indeed, \eqref{eq:subjective_probability} suggests that a possibility function can be seen as inducing an upper bound for probability distributions. It follows that selecting the sampling distribution from the set of upper-bounded probability distributions is also meaningful. A particular choice that is appropriate in many settings is to follow the maximum-entropy principle \cite{Jaynes1957} and consider the maximum-entropy distribution that is upper bounded by $f_{\bm{x}}$ as in \eqref{eq:subjective_probability}, as proposed in \cite{Houssineau2017_sequential}. When $\mathsf{X}$ is discrete, it is possible to further increase the entropy by replacing the set-wise upper bound of \eqref{eq:subjective_probability} by a point-wise upper bound of the form $p(x) \leq f_{\bm{x}}(x)$, $x \in \mathsf{X}$, with $p$ a probability mass function on $\mathsf{X}$. This approach will be particularly useful in the context of multi-object inference since it will lead to an increase of the diversity of explored data associations when compared to sampling from the distribution proportional to $f_{\bm{x}}$.

\subsection{Problem formulation}

The objective in the remainder of this section is to design a proposal distribution for identifying the mode of the possibility function $\hat{\Pi}$ defined in \eqref{eq:smoothingPath} via the Metropolis-Hastings algorithm. We assume for the moment that this proposal distribution is given and express it as a Markov kernel $\Phi$ from $\mathcal{A}$ to itself. A natural starting point for exploring the set $\mathcal{A}$ is to consider the case where all observations are false alarms, that is, we start from the element $A = \emptyset \in \mathcal{A}$. We first assume that $\hat{\Pi}$ can be evaluated everywhere so that, given a previous sample $A$, a new sample $A'$ can be obtained from the probability distribution $\Phi(\cdot \given A)$ and accepted with probability
\begin{equation}
\label{eq:acceptance_proba}
\hat{\alpha}_t(A,A') = \min\bigg(1, \dfrac{\hat{\Pi}(A')^{\rho_t}\Phi(A \given A')}{\hat{\Pi}(A)^{\rho_t}\Phi(A' \given A)} \bigg),
\end{equation}
where $t$ is the current iteration and $\rho_t$ is the inverse temperature defined by $\rho_0 = 1$ and $\rho_t = \rho_{t-1} / (1 - c)$ for some constant $c$.

The main difficulty with the Metropolis-Hastings algorithm in the context of interest is to design a proposal distribution $\Phi$ with adequate properties. In particular, there are two issues with this approach which we will aim to solve in the remainder of this section:
\begin{enumerate}[label=\roman*)]
\item The possibility function $\hat{\Pi}$ on $\mathcal{A}$ is highly multimodal in general so that moves that are local both in space and time are unlikely to yield a sufficient exploration of the space.
\item Implementing moves on entire paths in the set $\mathcal{O}_K$ would be more global in nature; however this requires the non-trivial introduction of additional structure on this set.
\end{enumerate}
These two issues will be addressed in Sections~\ref{sec:hisp} and \ref{sec:structure} respectively. Section~\ref{sec:proposalDataAssociation} will then detail the construction of the proposal distribution $\Phi$. Extensions of the MCMC algorithm introduced for $\hat{\Pi}$ to the possibility function $\Pi$ on $\mathcal{T}$ will be covered in Section~\ref{sec:trackMCMC}.

\subsection{Approximate multi-object filtering}
\label{sec:hisp}

\begin{figure}
\centering
    \begin{subfigure}[b]{0.49\textwidth}
        \centering
        \includegraphics[width=\textwidth, trim=25pt 75pt 20pt 65pt, clip]{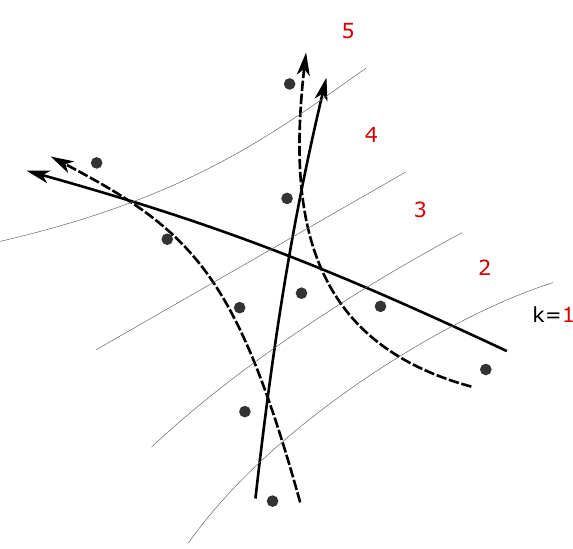}
        \caption{Reassignment of two paths when trajectories cross.}
        \label{fig:reassign2}
    \end{subfigure}
    \begin{subfigure}[b]{0.49\textwidth}
        \centering
        \includegraphics[width=\textwidth, trim=25pt 75pt 20pt 65pt, clip]{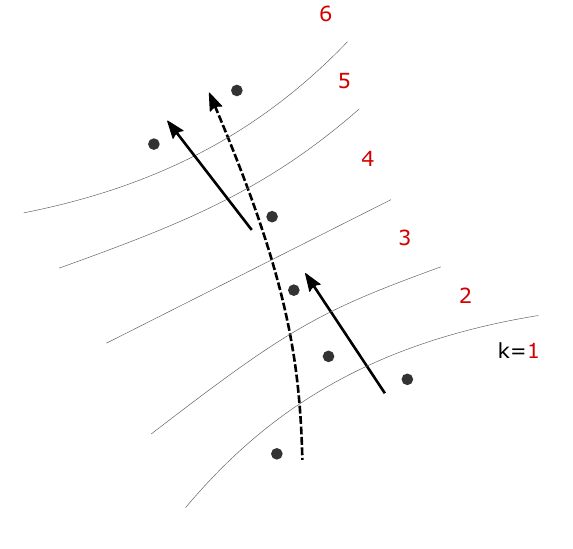}
        \caption{Track fragmentation and change of observation.}
        \label{fig:reassign1}
    \end{subfigure}
\caption{Examples of reassignment where black dots represent observations and different line styles represent different possible data associations. Thin grey lines and red numbers show the time steps to which different observations belong.}
\label{fig:reassign}
\end{figure}

In order to explore the different possible associations in the set $\mathcal{T}$ without getting stuck in local maxima and without incurring detrimental effects on the mixing of the MCMC chain, we propose to use a multi-object filter to ensure that any proposed association is meaningful from the viewpoint of the model. The motivation for leveraging the capabilities of an approximate filtering algorithm to solve the corresponding smoothing problem is very similar to the one behind particle MCMC \cite{Andrieu2010}. To illustrate the challenge with proposing changes in data association, we consider the case where two objects have crossing trajectories as in Figure~\ref{fig:reassign2}; if we only change one observation of a given path at a time, then it will take many moves to go from one high-credibility data association to another, and some of these moves will be in regions of arbitrarily small credibility. Alternatively, as is usual with MCMC algorithms, proposing bigger moves without taking into account the geometry of the target possibility function will result in an extremely low acceptance rate. This would be the case for instance if we were to reassign paths by simply proposing new observations uniformly at random. The objective is therefore to obtain paths that are consistent with the model given a restricted number of initial observations (first observation in a path). The corresponding moves that we will construct will be global in the sense that they might affect all time steps but local in sense that only a restricted number of paths will be (re)assigned. The considered filtering algorithm should have a low complexity in order to limit the computational cost of the overall MCMC algorithm. A possible candidate could therefore be the probability hypothesis density (PHD) filter \cite{Mahler2003} or its analogue in the context of possibility theory \cite{Houssineau2018_PHD}. However, the PHD filter does not solve the data association problem and, as q consequence, cannot be used to propose paths. Instead, we consider an analogue of the hypothesised filter for stochastic populations \cite{Houssineau2018_TSP}, or HISP filter, which is of the same complexity as the PHD filter and which allows for distinguishing objects.

At time step $k \in \{1,\dots,K\}$, the HISP filter provides the marginal probabilities for extending an existing path $z_{1:k-1} \in \bar{Z}_1 \times \dots \times \bar{Z}_{k-1}$ with an additional observation $z_k \in \bar{Z}_k$ at the current time step. The standard version of the algorithm would consider all such associations (at least the ones that are not too unlikely) and proceed to the next time step; however, we consider a modified version where a feasible data association is drawn at every time step so that the number of considered paths does not increase exponentially and the computational cost is further reduced. We also use the modelling based on possibility functions introduced in the previous sections instead of the probabilistic modelling considered in \cite{Houssineau2018_TSP}. The different steps of this modified HISP filter are given in the following sections.

The context is as follows: since only part of the existing paths are reassigned and since observations can only be associated with one object, it follows that some of the observations are unavailable to the HISP filter; we denote by $Z_k^-$ the sets of available observations at any time step $k \in \{1,\dots,K\}$. We will assume in this section that the credibility $\alpha_{\mathrm{nd}}$ of detection failure and the credibility $\alpha_{\mathrm{ns}}$ of non-survival are constant over the state space for the sake of simplicity; as opposed to the probabilistic case, this can be achieved in general by selecting the (constant) credibility of detection failure to be $\sup_{x \in \mathsf{X}} \alpha_{\mathrm{nd}}(x)$ and similarly for the credibility of non-survival. This operation can be seen as a voluntary loss of information with the purpose of gaining a property of interest.

\subsubsection{Initialisation}

We assume that, using local moves, the MCMC algorithm provides a set $Z_{\mathrm{c}} = \{(z^i_{\mathrm{c}},k^i_{\mathrm{c}})\}_{i=1}^{N_{\mathrm{c}}}$ of $N_{\mathrm{c}}$ pairs with $z^i_{\mathrm{c}}$ the initial observation for a path and with $k^i_{\mathrm{c}}$ the corresponding time step, $i \in \{1,\dots,N_{\mathrm{c}}\}$. These observations might or might not be at the same time step but the pairs $(z^i_{\mathrm{c}},k^i_{\mathrm{c}})$, $i \in \{1,\dots,N_{\mathrm{c}}\}$, are assumed to be different from each other. A path will be initialised every time one of these observations is encountered in the provided sets of observation $Z_k^-$.

\subsubsection{Prediction}

We denote by $O_{k-1}$ the set of paths at time $k-1$, that is the subset of $\mathcal{O}_{k-1} = \bar{Z}_1 \times \dots \times \bar{Z}_{k-1} \setminus \{\phi\}^{k-1}$ composed of paths that have been selected so far as potential sequences of object-originated observations. To each path $\op \in O_{k-1}$ corresponds a possibility function $f_{k-1}(\cdot \given \op)$ on the state space $\mathsf{X}$. Recalling that $g_k$ is the Markov transition from $\mathsf{X}$ to itself describing the objects' dynamics, we obtain the predicted possibility function
\begin{equation*}
f_{k|k-1}(x \given \op) = \sup_{x' \in \mathsf{X}} g_k( x \given x' ) f_{k-1}(x' \given \op), \qquad x \in \mathsf{X}.
\end{equation*}
Such a prediction only considers the event where the object survives to the $k$-th time step although it is possible for objects to disappear. We postpone considerations of this aspect of the prediction to a further stage in the algorithm.

\subsubsection{Update}

At time step $k$, the set of observations $Z_k^-$ is available to update the existing paths. For any path $\op$ in the set $O_{k-1}$ of previously selected paths and for any new observation $z \in Z_k^- \cup \{\phi\}$, the posterior possibility function associated with the extended path $\op:z$, with ``$:$'' denoting concatenation, is defined as
\begin{equation*}
f_k(x \given \op:z) =
\begin{cases*}
\dfrac{\ell_k(z \given x) f_{k|k-1}(x \given \op)}{\sup_{x' \in \mathsf{X}} \ell_k(z \given x') f_{k|k-1}(x' \given \op)} & if $z \in Z^-_k$ \\
f_{k|k-1}(x \given \op) & if $z = \phi$.
\end{cases*}
\end{equation*}
We can then select which observation in $Z_k^- \cup \{\phi\}$ will be used to propagate the path based on the credibility of the corresponding association. However, before expressing the latter, we first have to introduce the prior credibility of presence, which depends on the consecutive number of time steps for which the path under consideration has not been detected. Indeed, there is some remaining ambiguity whenever the empty observation $\phi$ is selected for a path since it is unclear in this case whether the detection has failed for the corresponding object or the object has not survived the last prediction step. We purposefully maintain this ambiguity and postpone the decision in order to better estimate which of these two events occur. Indeed, the credibility of non-survival is most often much lower than the credibility of detection failure, e.g.\ $\alpha_{\mathrm{nd}} = 0.1$ and $\alpha_{\mathrm{ns}} = 0.001$, so that terminating a track after a single detection failure is unlikely. Yet, if detection failures keep occurring for $l$ time steps, then the credibility of the corresponding events, i.e.\ $\alpha_{\mathrm{nd}}^l$ for the case where the object remains and $\alpha_{\mathrm{ns}}$ for the case where the object has disappeared, will rapidly favour a disappearance as opposed to a sequence of detection failures. For any path $\op \in O_{k-1}$, we denote by $l_{\op}$ the number of consecutive time steps for which $\phi$ has been selected, e.g.\ if $\op$ is of the form $(\op_1,\dots,\op_{k-3},\phi,\phi)$ with $\op_{k-3} \neq \phi$ then $l_{\op} = 2$. We then compute the credibility that the corresponding object has survived/not survived since the last detection as
\begin{equation*}
\hat{\alpha}_{\mathrm{s}}(\op) = \dfrac{\alpha_{\mathrm{nd}}^{l_{\op}}}{ \alpha_{\mathrm{ns}} \lor \alpha_{\mathrm{nd}}^{l_{\op}}},
\qquad
\hat{\alpha}_{\mathrm{ns}}(\op) = \dfrac{\alpha_{\mathrm{ns}}}{ \alpha_{\mathrm{ns}} \lor \alpha_{\mathrm{nd}}^{l_{\op}}},
\end{equation*}
with $a \lor b \doteq \max\{a,b\}$ for any $a,b \in \mathbb{R}$. The binary operator $\lor$ is assumed to have lower precedence than multiplication, so that $a \lor bc = a \lor (bc)$ for any $a,b,c \in \mathbb{R}$.

We can now express the marginal credibility of association on $Z_k^- \cup \{\phi\}$ for the path $\op \in O_{k-1}$ as
\begin{equation*}
\gamma_k(z \given \op) \propto
\Gamma_k\big(Z^-_k \setminus \{z\} \given O_{k-1} \setminus \op\big) L_k(z \given \op) 
\end{equation*}
for any observation $z \in Z^-_k \cup \{\phi\}$, with
\begin{equation*}
L_k(z \given \op) =
\begin{cases*}
\hat{\alpha}_{\mathrm{s}}(\op) \sup_{x \in \mathsf{X}} \ell_k(z \given x) f_{k|k-1}(x \given \op) & if $z \in Z^-_k$ \\
\hat{\alpha}_{\mathrm{ns}}(\op) \lor \hat{\alpha}_{\mathrm{s}}(\op)\big( \alpha_{\mathrm{ns}} \lor \alpha_{\mathrm{nd}} \big) & otherwise
\end{cases*}
\end{equation*}
the marginal likelihood for the observation $z$ and with $\Gamma_k(Z \given O)$ the credibility for paths in the set $O \subseteq O_{k-1}$ to be associated with observations in the set $Z \subseteq Z^-_k$, which can be expressed as
\begin{equation*}
\Gamma_k(Z \given O) = \max_{\sigma : O \to Z' \cup \{\phi\}} f_{\mathrm{fa}}(Z \setminus \sigma(O))  \prod_{\op \in O} L_k(\sigma(\op) \given \op),
\end{equation*}
where the maximum is over all mappings $\sigma$ from $O$ to $Z' \cup \{\phi\}$ that are injective on $Z'$ and where $\sigma(O)$ is the image of $O$ by $\sigma$, i.e. $\sigma(O) = \{ \sigma(\op) : \op \in O \}$. Although, the number of simultaneously reassigned paths will be limited in the context of interest, the number of observations in $Z$ can be extremely large so that the computation of $\Gamma_k(Z \given O)$ can be challenging. Yet, it is possible to rewrite this term by assuming that any two paths in $O$ are unlikely to obtain large marginal likelihoods from a single observation in $Z$, that is,  for any $\op, \op' \in O$ such that $\op \neq \op'$ and any $z \in Z$, there exists $z' \in Z$ such that
\begin{equation*}
L_k(z \given \op) L_k(z \given \op') \leq L_k(z \given \op) L_k(z' \given \op').
\end{equation*}
In the probabilistic version of this assumption \cite{Houssineau2018_TSP}, the left hand side needs to be equal to $0$, which is more constraining. It follows that $\Gamma_k(Z \given O)$ can be expressed as
\begin{equation*}
\Gamma_k(Z \given O) = f_{\mathrm{fa}}(Z) \prod_{\op \in O} \bigg[ L_k(\phi \given \op) \lor \max_{z \in Z} \dfrac{L_k(z \given \op)}{\alpha_{\mathrm{fa}}(z)} \bigg].
\end{equation*}
This result can be proved easily by developing the product in the approximated expression and removing the terms where a single observation is associated with several tracks. Using this expression, all the terms $\Gamma_k(Z^-_k \setminus \{z\} \given O_{k-1} \setminus \op)$, for any $z \in Z^-_k \cup \{\phi\}$ and any $\op \in O_{k-1}$, can be calculated with a computational complexity of order $|O_{k-1}||Z^-_k|$. The approach is similar to the one detailed in \cite{Houssineau2018_TSP} for the probabilistic case.

We then select an observation in $Z_k \cup \{\phi\}$ at random for each of the paths in $O_{k-1}$ using the marginal credibility of association $\gamma_k(\cdot\given\op)$ from the maximum entropy approach. There are two ways of enforcing the modelling assumption that paths cannot contain the same observation:
\begin{enumerate}[label=\roman*)]
\item \label{it:reject_sampling} use a rejection sampling strategy to ensure that only acceptable data associations are proposed, and
\item \label{it:just_reject} completely reject the proposed data association if it contains overlapping paths.
\end{enumerate}
The main drawback with the first option is that calculating the probability of proposing a given acceptable data association is combinatorial in nature and becomes a computational bottleneck when the number of observations is large. We therefore consider the second option.

Finally, we initialise a new path for any $(z^i_{\mathrm{c}}, k^i_{\mathrm{c}})$, $i \in \{1,\dots,N_{\mathrm{c}}\}$, such that $k^i_{\mathrm{c}} = k$. This path is of the form $\op = (\phi,\dots,\phi,z^i_{\mathrm{c}})$.

At the last time step, a decision is taken for all observations paths, even the one ending with empty observations, and a set $A_{\mathrm{c}}$ is defined as the set of all created paths. 
The conditional probability for generating the set of paths $A_{\mathrm{c}}$ given the initial observations $Z_{\mathrm{c}}$ and the available observations $Z^-_1, \dots, Z^-_K$ is denoted $P_{\mathrm{c}}(A_{\mathrm{c}} \given Z_{\mathrm{c}}, Z^-_{1:K})$.

\subsection{Structure on the set of paths}
\label{sec:structure}

In order to help exploring the set of data associations $\mathcal{A}$, it is useful to equip the underlying set of paths $\mathcal{O}_K$ with additional structure. The only natural structure on $\mathcal{O}_K$ is the one inherited from the fact that the observations are in the set $\mathsf{Z}$ which is a subset of an Euclidean space. This is not however sufficient since simply measuring the distance between two observations $z_k$ and $z'_{k'}$ at two different time steps $k \neq k'$ as $\| z_k - z'_{k'} \|$, with $\|\cdot\|$ the Euclidean norm, does not take into account the structure of the problem. Moreover, the notion of distance is very model-dependent and what is considered as ``close'' or ``far'' would need to be adjusted for each scenario. Instead, we use the objects' dynamical and observation model as a reference and relate observations via the credibility for these observations to be generated by the same object. These observations can be seen as \emph{consistent} if that credibility is close to $1$ and \emph{inconsistent} if it is close to $0$. In order to simplify the calculations, we assume the existence of an upper bounding function $g$ for the Markov transition $g_k$ such that
\begin{equation}
\label{eq:upperBound}
g_k(x \given x') \leq g(x \given x'), \qquad x,x' \in \mathsf{X},
\end{equation}
for any $k \in \{1,\dots,K\}$, with $g$ of the form
$$
g( x \given x') = \overline{\mathrm{N}}(x; Fx', Q), \qquad x,x' \in \mathsf{X},
$$
for some $d_{\mathsf{X}} \times d_{\mathsf{X}}$ matrices $F$ and $Q$. 

We consider two time steps $k,k' \in \{1,\dots,K\}$ such that $k < k'$ as well as two observations $z$ and $z'$ at time steps $k$ and $k'$ respectively and introduce $f_{k,k'}(z' \given z)$ as the possibility for an object initialised from $z$ at time step $k$ to be observed again at time step $k'$ at $z'$ in the absence of any other observation, that is
\begin{equation}
\label{eq:credZPrimeGivenZ}
f_{k'|k}(z' \given z) = \sup_{x,x' \in \mathsf{X}} \ell_k(z' \given x') g^l(x' \given x) f_k(x \given z)
\end{equation}
where $l = k' - k$, where $g^l$ is the $l$-th fold convolution of the transition $g$, that is
\begin{equation}
\label{eq:lFoldCompG}
g^l(x_{k'} \given x_k) = \sup_{x_{k+1},\dots,x_{k'-1} \in \mathsf{X}} g(x_{k'} \given x_{k'-1}) \dots g(x_{k+1} \given x_k), \qquad x_k, x_{k'} \in \mathsf{X},
\end{equation}
and where $f_k(\cdot \given z)$ is the posterior possibility function defined as
$$
f_k(x \given z) = \dfrac{\ell_k(z \given x) f_0(x)}{\sup_{x' \in \mathsf{X}} \ell_k(z \given x') f_0(x')}, \qquad x \in \mathsf{X}.
$$
The possibility function $g^l(\cdot\given x_k)$ is an upper bound for the convolution of the Markov transitions $g_{k+1}, \dots, g_{k'}$. Assuming that $f_k(\cdot \given z) = \overline{\mathrm{N}}(m_z, \Sigma_0)$ and denoting by $\Sigma_l$ the covariance matrix after $l$ predictions, e.g.\ $\Sigma_1 = F\Sigma_0F^{\tr} + Q$, then the possibility function $f_{k'|k}(\cdot \given z)$ can be written
\begin{equation*}
f_{k'|k}(z' \given z) = \overline{\mathrm{N}}( z';  H_{k',z,l} F^l m_z,  H_{k',z,l}\Sigma_l H_{k',z,l}^{\tr} + R_{k'})
\end{equation*}
where $ H_{k',z,l}$ is the Jacobian of $h_{k'}$ at the point $F^lm_z$; the value of $d_{k,k'}(z, z')$ can be easily deduced. 

The main drawback of this notion of consistency is that observations tend to become more consistent as $l$ increases since there is more uncertainty about the state of the object as time passes by. To address this potential issue, we take the credibility of detection $\alpha_{\mathrm{d}}(\cdot)$ into account and focus on the credibility for an observation $z'$ to be the \emph{next} observation of the object after $z$. To fit into the considered context, we introduce a lower bound $a_{\mathrm{nd}}$ for the credibility of non-detection, i.e.\ $a_{\mathrm{nd}}$ is such that $\alpha_{\mathrm{nd}}(x) \geq a_{\mathrm{nd}}$ for any $x \in \mathsf{X}$. It then follows that the possibility for $z'$ to be the next observation after $z$ is
\begin{equation*}
\hat{f}_{k'|k}(z' \given z) = a_{\mathrm{nd}}^{l-1} f_{k'|k}(z' \given z),
\end{equation*}
defined for any $l > 0$. The function $\hat{f}_{k,k'}$ can be easily extended to $\mathsf{Z} \cup \{\phi\}$ by defining $\hat{f}_{k,k'}(\cdot \given \phi) = \hat{f}_{k,k'}(\phi \given \cdot) = 0$. 

\begin{figure}
\centering
    \begin{subfigure}[b]{0.49\textwidth}
        \includegraphics[width=\textwidth, trim=110pt 265pt 120pt 285pt, clip]{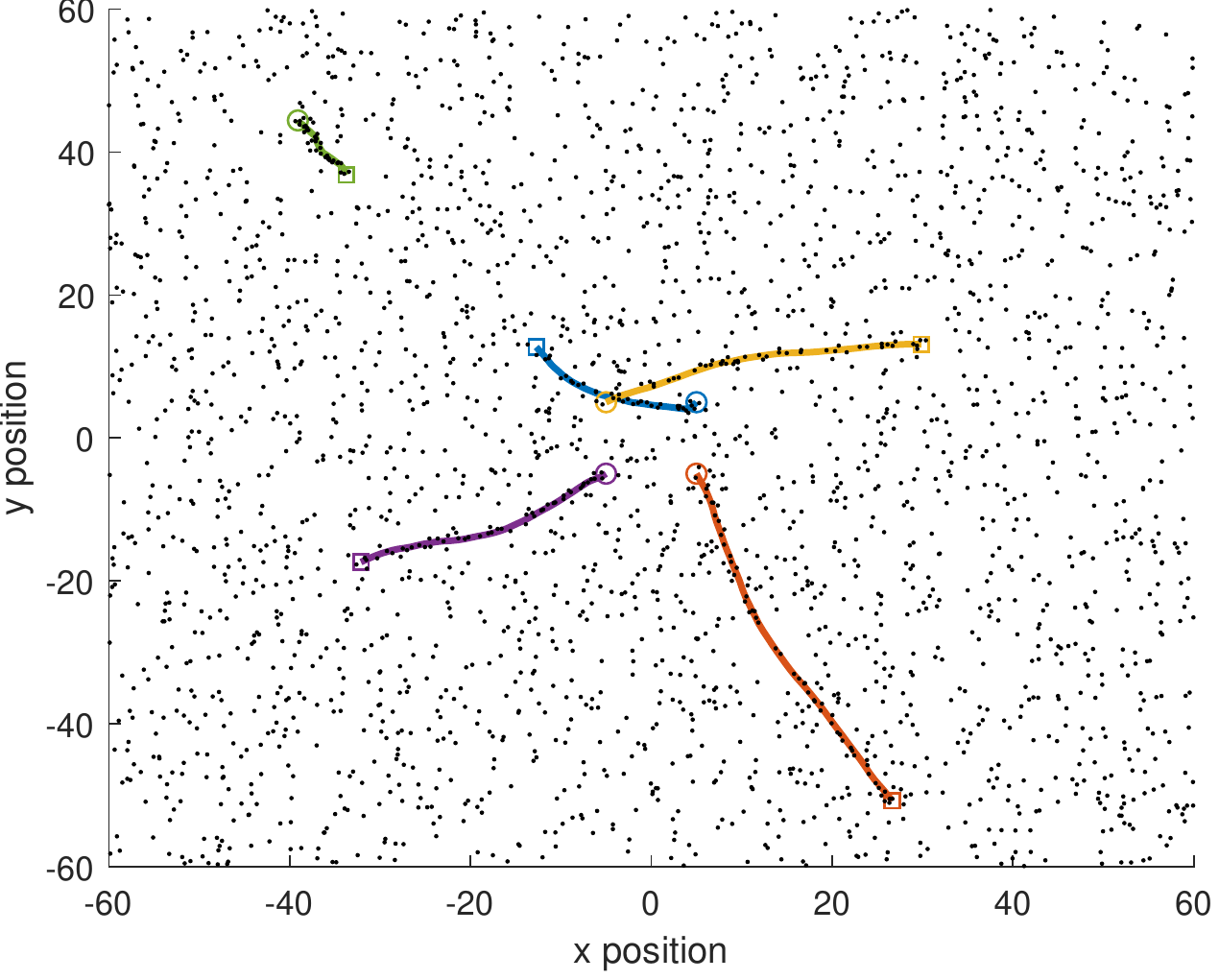}
        \caption{True trajectories as coloured lines, initial/final positions as circles/squares and observations as black dots.}
        \label{fig:scenario_scatter}
    \end{subfigure}
    \begin{subfigure}[b]{0.49\textwidth}
        \includegraphics[width=\textwidth, trim=110pt 265pt 120pt 285pt, clip]{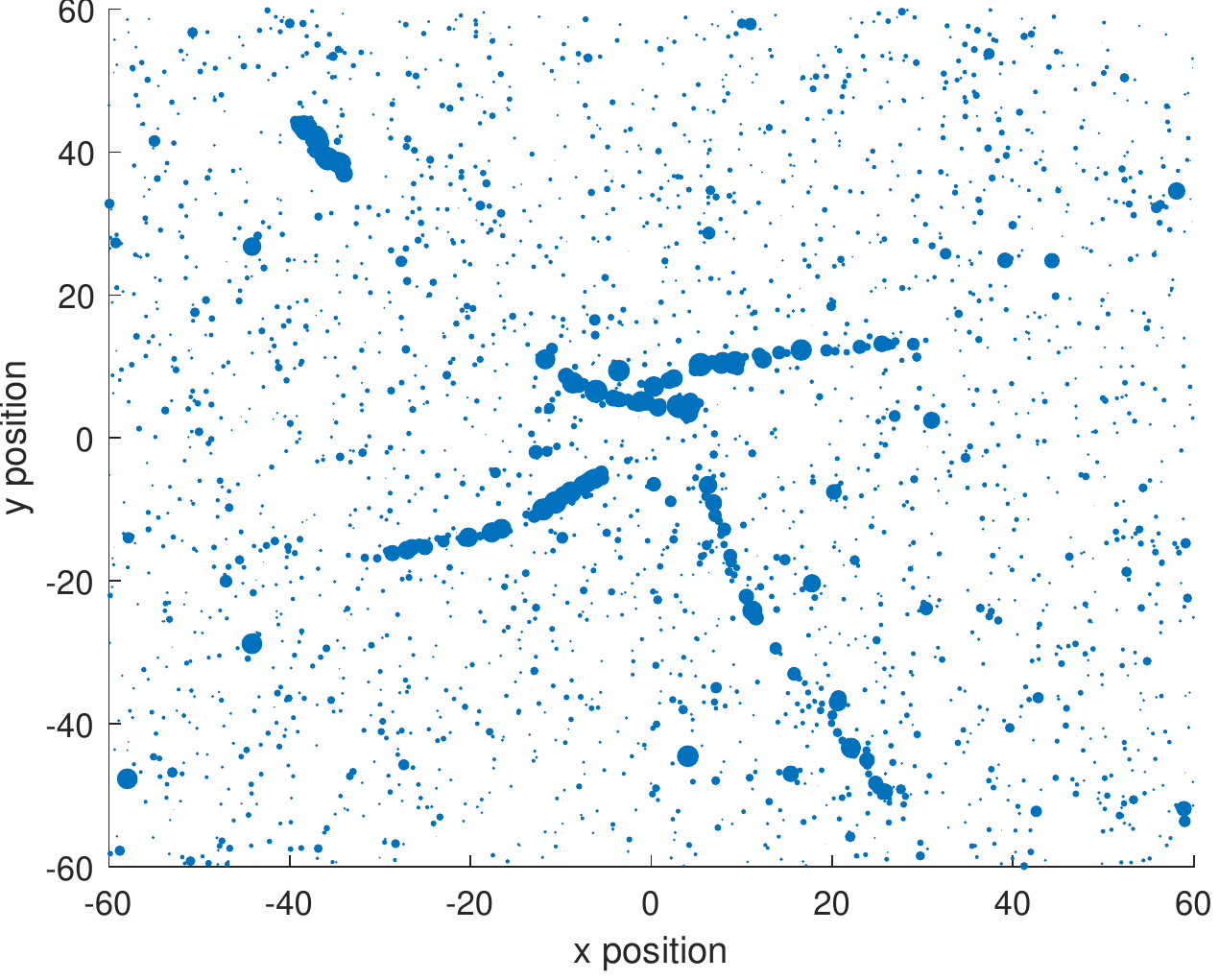}
        \caption{Scatter plot of the observations with a size proportional to the corresponding credibilities of selection.}
        \label{fig:scatter}
    \end{subfigure}
\caption{Scenario with $6$ objects as represented in (A) with the corresponding scatter plot of the probability for ach observation to be selected in (B).}
\label{fig:scatter_obs_dist}
\end{figure}

\begin{example}
To illustrate the use of the notion $\hat{f}_{k,k'}$ of consistency, a simple scenario consisting of $6$ objects is considered as in Figure~\ref{fig:scenario_scatter}. For each observation $z \in Z_k$ at some time step $k \in \{1,\dots,K\}$ we compute a marginal credibility for $z$ as
\begin{equation}
\label{eq:dist2nextObs}
\hat{f}_k(z) = \max_{k' : k' > k} \Big( \max_{z' \in Z_{k'}} \hat{f}_{k',k}(z' \given z) \Big).
\end{equation}
The scalar $\hat{f}_k(z)$ can be interpreted as the credibility for $z$ to be followed by another observation in $Z_{k'}$ for some $k' > k$. When creating a new track, we can then define the probability for selecting $z$ as the first observation of the new track as a function of a $\hat{f}_k(z)$. A scatter plot displaying these credibilities for all observations is shown in Figure~\ref{fig:scatter}. 
\end{example}

The advantage of relating observations in this way is that it can be easily extended to paths. Indeed, we can define the consistency between an observation $z \in Z_k$ at some given time $k$ with a path $\op$ in $\mathcal{O}_K$ as
\begin{equation}
\label{eq:dist2track}
\hat{f}_k(z, \op) = \max\Big\{ \max_{k' : k' < k} \hat{f}_{k,k'}( z \given \op_{k'}),\, \max_{k' : k' > k} \hat{f}_{k',k}( \op_{k'} \given z) \Big\},
\end{equation}
where the initial observation is either $z$ or one of the observations in $\op$. Similarly, the consistency between two paths $\op$ and $\op'$ in $\mathcal{O}_K$ is defined as
$$
\hat{f}(\op, \op') = \max\Big\{ \max_{k,k' : k' < k} \hat{f}_{k,k'}( \op_k \given \op'_{k'}),\, \max_{k,k' : k' > k} \hat{f}_{k',k}( \op'_{k'} \given \op_k) \Big\}
$$
We can now propose to modify a given data association by changing nearby paths, and therefore focus the computational power on moves that are likely to be accepted. Although the approach considered here is not standard, it has two main advantages: it relates observations together and applies to non-linear cases as long as a Gaussian upper-bounding function can be found.

In practice, it might be necessary to reduce the time required for computing $\hat{f}_{k',k}(z' \given z)$ between any pair $(z,z')$ of observations, especially if the scenario runs over many times steps or if the number of observations at every time step is large. In that case, one can define a threshold $\tau'$ such that if $a_{\mathrm{nd}}^l < \tau'$ then any observations that are $l$ time steps apart will be arbitrarily assigned a credibility of $0$.

\subsection{Design of the proposal distribution}
\label{sec:proposalDataAssociation}

When designing a proposal distribution $\Phi$ for our MCMC algorithm, several requirements need to be considered: it should be possible to
\begin{enumerate}[label=\roman*)]
\item \label{it:reassign_multiple} reassign several paths simultaneously in order to address crossings as illustrated in Figure~\ref{fig:reassign2} and track fragmentation,
\item \label{it:reassign_init} reassign both the initial observation of a path and the subsequent path as in Figure~\ref{fig:reassign1}, and
\item \label{it:create_track} create a new path.
\end{enumerate}
Requirement~\ref{it:reassign_multiple} can be easily fulfilled by using the approach presented in Section~\ref{sec:hisp} however, instead of simply choosing the paths at random, it is more efficient to focus on nearby paths. In order to simultaneously reassign the initial observations of a given set of paths $A_{\mathrm{r}}$ as needed in Requirement~\ref{it:reassign_init}, we consider the notion of consistency defined in \eqref{eq:dist2track}. Once a new initial observation has been selected, the approach of Section~\ref{sec:hisp} can be used to reassign the rest of the chosen path. Finally, the marginal consistency defined in \eqref{eq:dist2nextObs} can be used for Requirement~\ref{it:create_track} in order to identify observations that are likely to be initial observations. 

The general objective is to find a proposal distribution $\Phi$ that is as simple as possible and such that the associated MCMC kernel is irreducible and reversible. Starting from a given set of paths $A$ of size $s = |A|$, we suggest to proceed as follows:
\begin{enumerate}[label=\arabic*),wide]
\item Sample a number $N_{\mathrm{r}}$ of paths to reassign from a probability mass function (p.m.f.) $p_{\mathrm{r}}(\cdot \given s)$ such that $N_{\mathrm{r}} \leq s$ almost surely (a.s.), e.g.\ a truncated Poisson distribution. Then, sample the number $N_{\mathrm{c}}$ of paths to be created from the p.m.f.\ $p_{\mathrm{c}}(\cdot \given N_{\mathrm{r}})$ on the set of non-negative integers $\mathbb{N}$ defined as
\begin{equation*}
p_{\mathrm{c}}(n \given N_{\mathrm{r}}) 
\begin{cases*}
\delta_1(n) & if $N_{\mathrm{r}} = 0$ \\
\tilde{p}_{\mathrm{c}}(n - N_{\mathrm{r}}) & otherwise,
\end{cases*}
\end{equation*}
with $\tilde{p}_{\mathrm{c}}$ a p.m.f.\ on $\{-1,0,1\}$ to be defined. With this model, there will be one created path a.s.\ when none are reassigned (there is limited interest in creating several paths at once in this case) and the number of paths will be increased by one, kept constant or decreased by one in case of reassignment. Reducing the number of paths by one will address the issue of track fragmentation, keeping the number of paths constant is appropriate when considering objects with crossing trajectories, and leaving the possibility of increasing the number of paths is required to ensure reversibility. Indeed, when evaluating the probability of the reverse proposal, created paths will become reassigned paths and vice versa.
\item If $N_{\mathrm{r}} = 0$ then define $A_{\mathrm{r}} = \emptyset$ and proceed to the next step, otherwise, select the set $A_{\mathrm{r}} = \{\op_i\}_{i=1}^{N_{\mathrm{r}}}$ of paths to be reassigned as follows: the first path $\op_1$ is picked uniformly at random from the set of paths $A$ then the $N_{\mathrm{r}}-1$ remaining paths, if any, are selected based on their distance to $\op_1$:
\begin{equation*}
\op_i \sim \mathcal{P}_{\op_{1:i-1}}\big( \hat{f}(\op_1,\cdot) \big)
\end{equation*}
for any $1 < i \leq N_{\mathrm{r}}$, where $\mathcal{P}_{\op_{1:i-1}}(\cdot)$ is a function transforming possibility functions into probability distributions, e.g.\ the maximum-entropy distribution upper-bounded point-wise by $\hat{f}(\op_1,\cdot)$, which we assume to verify
\begin{equation*}
\sum_{\op \in A} \mathcal{P}_{\op_{1:i-1}}\big( \hat{f}(\op_1,\cdot)\big)(\op) = 1 \AND \mathcal{P}_{\op_{1:i-1}}\big(\hat{f}(\op_1,\cdot)\big)(\op_j) = 0, \quad j \in \{1,\dots,i-1\}.
\end{equation*}
Therefore, the set of paths $A_{\mathrm{r}}$ is sampled without replacement from the set $A$. When evaluating the probability $P_{\mathrm{r}}(A_{\mathrm{r}} \given N_{\mathrm{r}})$ for sampling the subset $A_{\mathrm{r}}$ of $A$, all possible ways of obtaining such a subset must be taken into account, that is
\begin{equation*}
P_{\mathrm{r}}(A_{\mathrm{r}} \given N_{\mathrm{r}}, A) =
\begin{dcases*}
\dfrac{1}{s} \sum_{\sigma \in \mathrm{Sym}(N_{\mathrm{r}})} \prod_{i=2}^{N_{\mathrm{r}}} \mathcal{P}_{(\op_{\sigma(1)},\dots,\op_{\sigma(i-1)})}\big(\hat{f}(\op_1,\cdot)\big)(\op_{\sigma(i)}) & if $N_{\mathrm{r}} > 0$ \\
\delta_{\emptyset}(A_{\mathrm{r}}) & otherwise,
\end{dcases*}
\end{equation*}
where $\mathrm{Sym}(n)$ is the set of permutations of $\{1,\dots,n\}$. Although the computational complexity for this term is combinatorial, $N_{\mathrm{r}}$ is usually small so the actual computational time is limited. 
\item If $N_{\mathrm{c}} = 0$ then define $Z_{\mathrm{c}} = \emptyset$ and proceed to the next step, otherwise, select the $N_{\mathrm{c}}$ initial observations $Z_{\mathrm{c}} = \{(z^i_{\mathrm{c}},k^i_{\mathrm{c}})\}_{i=1}^{N_{\mathrm{c}}}$ from the set $\bigcup_{k=1}^K \{(z,k) : z \in Z^-_k\}$ of available observations, with $Z^-_k$ defined for any $k \in \{1,\dots,K\}$ as
\begin{equation*}
Z^-_k = \big\{ z \in Z_k: \forall (\op,k_-) \in A \setminus A_{\mathrm{r}}, z \neq \op_k \big\}.
\end{equation*}
The selection of the initial observations is performed without replacement as
\begin{equation*}
\hat{z}^i_{\mathrm{c}} \sim \mathcal{P}_{(\hat{z}^1_{\mathrm{c}},\dots,\hat{z}^{i-1}_{\mathrm{c}})}\big(\hat{f}^{N_{\mathrm{r}}}_{\mathrm{c}}\big)
\end{equation*}
where $\hat{z}^j_{\mathrm{c}}$ stands for the pair $(z^j_{\mathrm{c}}, k^j_{\mathrm{c}})$ for any $j \in \{1,\dots,N_{\mathrm{c}}\}$, and where the possibility function $\hat{f}^{N_{\mathrm{r}}}_{\mathrm{c}}$ is defined as the marginal consistency \eqref{eq:dist2nextObs} if $N_{\mathrm{r}} = 0$ and as the consistency \eqref{eq:dist2track} with the future observation in the paths in $A_{\mathrm{r}}$ otherwise. Indeed, when reassigning $N_{\mathrm{r}} > 0$ paths, it is more efficient to propose new paths in the same area rather than initialising paths in random locations, especially during the burn-in period of the MCMC when observations in different places are likely to originate from objects. The probability of proposing the subset $Z_{\mathrm{c}}$ of observations takes a similar form as for path reassignment and can be expressed as
\begin{equation*}
\tilde{P}_{\mathrm{c}}\big(Z_{\mathrm{c}} \given N_{\mathrm{c}}, Z^-_{1:K}\big) =
\begin{dcases*}
\sum_{\sigma \in \mathrm{Sym}(N_{\mathrm{c}})} \prod_{i=1}^{N_{\mathrm{c}}} \mathcal{P}_{(\hat{z}^{\sigma(1)}_{\mathrm{c}},\dots,\hat{z}^{\sigma(i-1)}_{\mathrm{c}})}\big(\hat{f}^{N_{\mathrm{r}}}_{\mathrm{c}}\big)(\hat{z}^{\sigma(i)}_{\mathrm{c}}) & if $N_{\mathrm{c}} > 0$ \\
\delta_{\emptyset}(Z_{\mathrm{c}}) & otherwise.
\end{dcases*}
\end{equation*}
The comment regarding computational complexity made about $P_{\mathrm{r}}(\cdot \given N_{\mathrm{r}})$ applies equally here.
\item Apply the approximate multi-object filter of Section~\ref{sec:hisp} to the set of initial observations $Z_{\mathrm{c}}$ and with the sets of available observations $Z^-_1,\dots,Z^-_K$ and denote $A_{\mathrm{c}}$ the generated set of paths. If $A_{\mathrm{c}} \cap A_{\mathrm{r}} \neq \emptyset$ then we reject the proposal, otherwise, the proposed set of paths is $A' = (A \setminus A_{\mathrm{r}}) \cup A_{\mathrm{c}}$. The reason for rejecting the proposal when $A_{\mathrm{c}} \cap A_{\mathrm{r}} \neq \emptyset$ is to ensure that $A_{\mathrm{c}}$ and $A_{\mathrm{r}}$ can be recovered from $A$ and $A'$ as $A_{\mathrm{c}} = A' \setminus A$ and $A_{\mathrm{r}} = A \setminus A'$.
\end{enumerate}

If the proposal has not been already rejected during its construction, the probability $\Phi(A' \given A)$ to go from the previous set of paths $A$ to the new set of paths $A'$ is computed as
\begin{equation*}
\Phi(A' \given A) = P_{\mathrm{c}}\big(A_{\mathrm{c}} \given Z_{\mathrm{c}}, Z^-_{1:K}\big) \tilde{P}_{\mathrm{c}}\big(Z_{\mathrm{c}} \given N_{\mathrm{c}}, Z^-_{1:K}\big) P_{\mathrm{r}}(A_{\mathrm{r}} \given N_{\mathrm{r}}, A) p_{\mathrm{c}}(N_{\mathrm{c}} \given N_{\mathrm{r}}) p_{\mathrm{r}}(N_{\mathrm{r}} \given s).
\end{equation*}
The probability $\hat{\alpha}(A,A')$ of accepting the proposed set of paths $A'$ can then be computed using \eqref{eq:acceptance_proba}.

\section{MCMC on the set of tracks}
\label{sec:trackMCMC}

We now want to design a MCMC algorithm that targets the possibility function $\Pi$ as introduced in \eqref{eq:smoothingTrack}. In this case, the Metropolis-Hastings acceptance ratio is
\begin{equation}
\label{eq:acceptance_proba2}
\alpha_t(T,T') = \min\bigg(1, \dfrac{\Pi(T')^{\rho_t} \Psi(T \given A)\Phi(A \given A')}{\Pi(T)^{\rho_t}\Psi(T' \given A')\Phi(A' \given A)} \bigg),
\end{equation}
with $A$ and $A'$ the set of paths in $T$ and $T'$ respectively. We therefore have to propose a time of appearance and a last time of existence for each path in $A$. These time steps will sampled independently from their previous values in $T$.

\subsection{Proposing the interval of existence}
\label{sec:proposalTime}

The objective in this section is to propose a time of appearance $m$ and a last time of existence $n$ for a given path $\op \in \mathcal{O}_K$, using the different quantities introduced in Section~\ref{sec:hisp}. We consider a path $\op \in \mathcal{O}_K$ of the form $\op':\phi$. One can sample the lag corresponding to the last time of appearance according to the probability mass function $p_-(\cdot \given \op)$ on $\{0,\dots,l_{\op}\}$ defined as the maximum-entropy distribution bounded by $l \mapsto \alpha_{\mathrm{ns}}^{\mathbbm{1}(l < l_{\op})}\alpha_{\mathrm{nd}}^l$. The last time of existence is set to $n = K - l_{\op} + L_-$. For the time of appearance $m$ associated with a path $\op \in \mathcal{O}_K$, we can simply sample a lag $L_+$ from the maximum-entropy distribution $p_+(\cdot \given \op)$ bounded by $l \mapsto \alpha_{\mathrm{nd}}^l$ and set $n = k_+(\op) - L_+$ with $k_+(\op)$ the time of the first observation in $\op$. The probability distribution $\Psi(\cdot \given A)$ is then associated with the proposal of a time of appearance and a last time of existence for each path in a given set $A \in \mathcal{A}$, i.e.\
$$
\Psi(T \given A) = \delta_{A}(\kappa(T))\prod_{(\op,m,n) \in T} \big[ p_+(m \given \op) p_-(n \given \op) \big].
$$

\subsection{Evaluating the marginal likelihood}
\label{sec:proposalTrajectory}

So far, the proposed approach does not assume a specific model for the dynamics and for the observation process. Indeed, although the likelihood $\ell_k(\cdot \given x)$ is assumed to take the form of a Gaussian possibility function, the function $h$ relating states to observations is general. We will however distinguish two different cases for the evaluation of the marginal likelihood: the linear-Gaussian case in which Kalman filtering can be used and the non-linear case where sequential Monte Carlo techniques are a natural alternative.

\subsubsection{Linear-Gaussian case} If the Markov transition $g_k$ is of the form $g_k(\cdot \given x') = \overline{\mathrm{N}}(F_k x', Q_k)$ for some $d_{\mathsf{X}} \times d_{\mathsf{X}}$ matrices $F_k$ and $Q_k$ and for any $k \in \{1,\dots,K\}$ and if the observation function $h_k$ is of the form $h_k(x) =  H_k x$ then the posterior distribution of the state at any time step can be computed analytically via the Kalman filter. In particular, for a given path $\op \in \mathcal{O}_{k-1}$, we denote by $m_k^{\op}$ and $\Sigma_k^{\op}$ the mean and variance of the state at time $k \in \{1,\dots,K\}$ given the observations in the path $\op$. The only difference with the standard Kalman filtering equation is the marginal likelihood which, due to the form of the likelihood, is expressed at time step $k$ as
$$
\hat{\ell}_k(z \given \op) = \sup_{x \in \mathsf{X}} \ell_k(z \given x) f_{k|k-1}(x \given \op) = \overline{\mathrm{N}}(z;  H_k m_k^{\op},  H_k \Sigma_k^{\op}  H_k^{\tr} + R_k)
$$
for any $z \in Z_k$.

\subsubsection{Non-linear case}

If either the objects' dynamics or the observation function is not linear, then there is no analytical form for the filtering distributions at different time steps in general. Sequential Monte Carlo (SMC) methods are an alternative to the Kalman filter in this case. An analogue \cite{Houssineau2017_sequential} of the bootstrap particle filter \cite{Gordon1993} can be used, see also \cite{Ristic2018, Ristic2019}. In particular, for a given path $\op \in \mathcal{O}_{k-1}$, we denote by $\{(w^{\op}_{k-1,i}, x^{\op}_{k-1,i})\}_{i=1}^N$ the indexed family of weighted particles approximating the predicted possibility function $f_{k|k-1}(\cdot \given \op)$, i.e.\
$$
\bar{\mathbb{E}}(\varphi(\bm{x}_{k}) \given \op) \approx \max_{1 \leq i \leq N} w^{\op}_{k-1,i} \varphi(x^{\op}_{k-1,i})
$$
for any real-valued function $\varphi$ on $\mathsf{X}$, with the uncertain variable $\bm{x}_k$ being described by $f_{k|k-1}(\cdot \given \op)$. Then
$$
\bar{\mathbb{E}}(\varphi(\bm{x}_{k}) \given \op:z) \approx \dfrac{\max_i w^{\op:z}_{k,i} \varphi(x^{\op}_{k-1,i})}{\max_i w^{\op:z}_{k,i}}
$$
for any $z \in Z_k$, where $w^{\op:z}_{k,i} =  w^{\op}_{k-1,i} \ell_k(z \given x^{\op}_{k-1,i})$ for any $i \in \{1,\dots,N\}$. In this situation, the marginal likelihood at time step $k$ can be approximated by
$$
\hat{\ell}_k(z \given \op) \approx \max_{1\leq i \leq N} w^{\op:z}_{k,i}.
$$

\section{Simulations}
\label{sec:simulations}

In all the cases to be considered, $K=50$ and $\mathsf{X} = \mathbb{R}^4$. States at time step $k$ are of the form $x_k = (\mathbf{x}_k, \mathbf{y}_k, \dot{\mathbf{x}}_k, \dot{\mathbf{y}}_k)^{\tr}$, where $\mathbf{x}_k$ and $\mathbf{y}_k$ are the coordinates of the position in the 2-dimensional Euclidean space and where $\dot{\mathbf{x}}_k$ and $\dot{\mathbf{y}}_k$ are the coordinates of the velocity. The duration of one time step is denoted $\Delta$ and the motion model is assumed to be of the form
$$
q_k(x_k \given x_{k-1}) = \mathcal{N}( x_k; F x_{k-1}, Q)
$$
with
$$
F = \begin{bmatrix}
1 & 0 & \Delta & 0 \\
0 & 1 & 0 & \Delta \\
0 & 0 & 1 & 0 \\
0 & 0 & 0 & 1
\end{bmatrix}
\AND
Q = \sigma_{\mathrm{a}}^2 \begin{bmatrix}
\Delta^4/4 & 0 & \Delta^3/2 & 0 \\
0 & \Delta^4/4 & 0 & \Delta^3/2 \\
\Delta^3/2 & 0 & \Delta^2 & 0 \\
0 & \Delta^3/2 & 0 & \Delta^2
\end{bmatrix},
$$
where $\sigma_{\mathrm{a}}$ is the standard deviation of the zero-mean random acceleration, which is considered as a noise term. This model is referred to as the nearly-constant velocity model. We will consider in particular the case where $\Delta = 1$ and $\sigma_{\mathrm{a}} = 0.05$.

For the sake of simplicity, the observation model is assumed to be linear; the position $(\mathbf{x}_k, \mathbf{y}_k)^{\tr}$ of an object is observed directly, which leads to $h(x_k) = Hx_k$ with
$$
H = \begin{bmatrix}
1 & 0 & 0 & 0 \\
0 & 1 & 0 & 0
\end{bmatrix}.
$$
The variance $R$ is of the form $\sigma^2 \bm{I}_2$ with $\sigma > 0$ and $\bm{I}_2$ the identity matrix of dimension $2$. This model is useful when tracking directly in the coordinate systems defined by a sensor such as the image plane of a camera. Other situations where this model arises are when multiple sensors provide complex observations which can be combined into a single observation before being used in a tracking algorithm such as with GPS or with multiple-input multiple-outputs sensor systems \cite{Bekkerman2006,Haimovich2007,Pailhas2016}. We will consider in particular the case where $\sigma = 0.3$ and $\mathsf{Y} = [-60, 60] \times [-60, 60]$.

\subsection{Parametrisation of the proposed algorithm}

If the probability of detection is $p_{\mathrm{d}}$ then the possibility of detection failure is set to $\alpha_{\mathrm{nd}} = 1 - p_{\mathrm{d}}$ and the possibility of detection $\alpha_{\mathrm{d}}$ is set to $1$. The same approach is used with the probability of survival. The possibility function $\alpha_{k,\mathrm{fa}}$ is assumed to be constant and equal to $10^{-2}$ for all scenarios; this is in spite of the fact that the number of false alarms will vary significantly across the considered settings. The reason for this is that $\alpha_{k,\mathrm{fa}}^n$ is seen as an upper bound for the probability of having $n$ false alarms. A similar approach is used for appearing objects with $f_{k,+}(n) = \alpha_+^n$ with $\alpha_+ = 10^{-4}$. The other model parameters such as $\sigma$ and $\sigma_{\mathrm{a}}$ are assumed to be known.

The proposed approach is compared to the MCMC for Data Association (MCMC-DA) method introduced in \cite{Oh2009}. In order to make the two methods comparable, the possibility function $\Pi$ is used to evaluate the log-likelihood of the proposed sets of tracks. However, as opposed to the proposed approach, MCMC-DA is provided with the true parameters of the model in the design of the corresponding proposal distribution.

\subsection{Choice of parameter}

We assume that the current sample from $\Pi$ is $T \in \mathcal{T}$ and denote by $A = \kappa(T)$ the corresponding set of paths. We then comment on the choice of parameters for the different steps in the proposal mechanism.

The number $N_{\mathrm{r}}$ of tracks to reassign is chosen from a Poisson distribution with parameter $\lambda_{\mathrm{r}} = 1$, truncated to the interval $\{0, \dots, |A|\}$. The parameter $\lambda_{\mathrm{r}}$ can be adjusted depending on the considered scenario: if objects are expected to be very close to each other and to frequently have crossing trajectories, then $\lambda_{\mathrm{r}}$ could be increased to raise the average number of tracks that are reassigned at once. Large reassignments are however less likely to be accepted so that a trade-off between exploration and mixing must be found, as is usual with MCMC.

The distribution $\tilde{p}_{\mathrm{c}}(\cdot \given N_{\mathrm{r}})$ on $\{-1, 0, 1\}$ drives the increase or decrease of the number of tracks in the proposal step. Since one of the main issues with the MCMC approach for data association is track fragmentation, i.e.\ the representation of a single object by a series of shorter tracks, it is generally helpful to focus on reducing the number of tracks. We therefore consider the following parametrisation:
$$
\tilde{p}_{\mathrm{c}}(\delta \given N_{\mathrm{r}}) =
\begin{cases*}
1/2 & if $\delta = -1$ \\
1/4 & if $\delta = 0$ \\
1/4 & if $\delta = 1$.
\end{cases*}
$$

\subsection{MCMC on the data association set}

The choice of parameter as well as the performance of the proposed approach are assessed on different scenarios.

\subsubsection{Simple scenario}

\begin{figure}
\centering
    \begin{subfigure}[b]{0.49\textwidth}
        \centering
        \includegraphics[width=\textwidth, trim=100pt 265pt 115pt 280pt, clip]{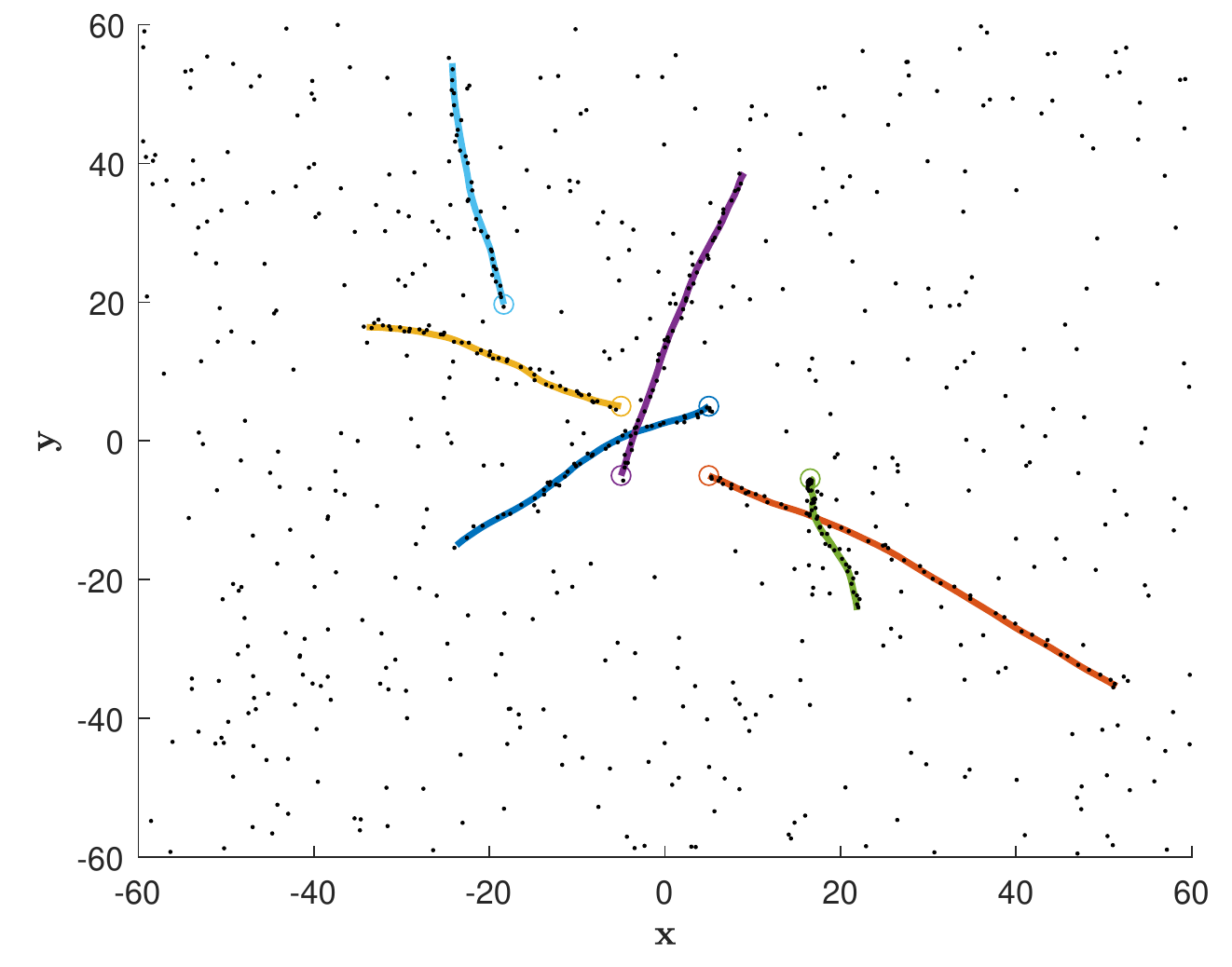}
        \caption{Trajectories (with circles indicating initial position) and observations (black dots).}
        \label{fig:scen_simple_pos}
    \end{subfigure}
    \begin{subfigure}[b]{0.49\textwidth}
        \centering
        \includegraphics[width=\textwidth, trim=100pt 265pt 115pt 280pt, clip]{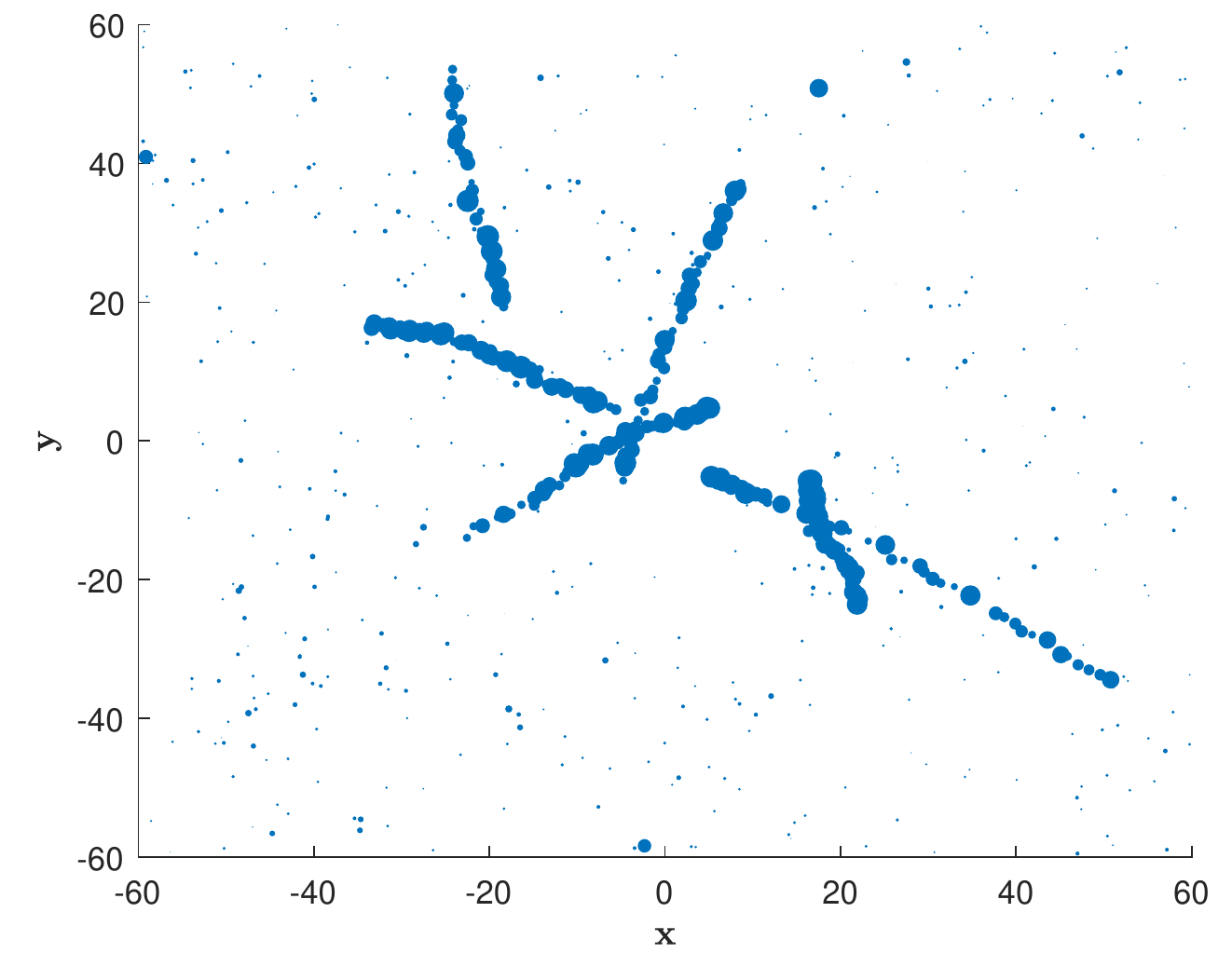}
        \caption{Distance from one given observation to the nearest observation.}
        \label{fig:scen_simple_dist}
    \end{subfigure}
    \begin{subfigure}[b]{0.49\textwidth}
        \centering
        \includegraphics[width=\textwidth, trim=95pt 265pt 115pt 280pt, clip]{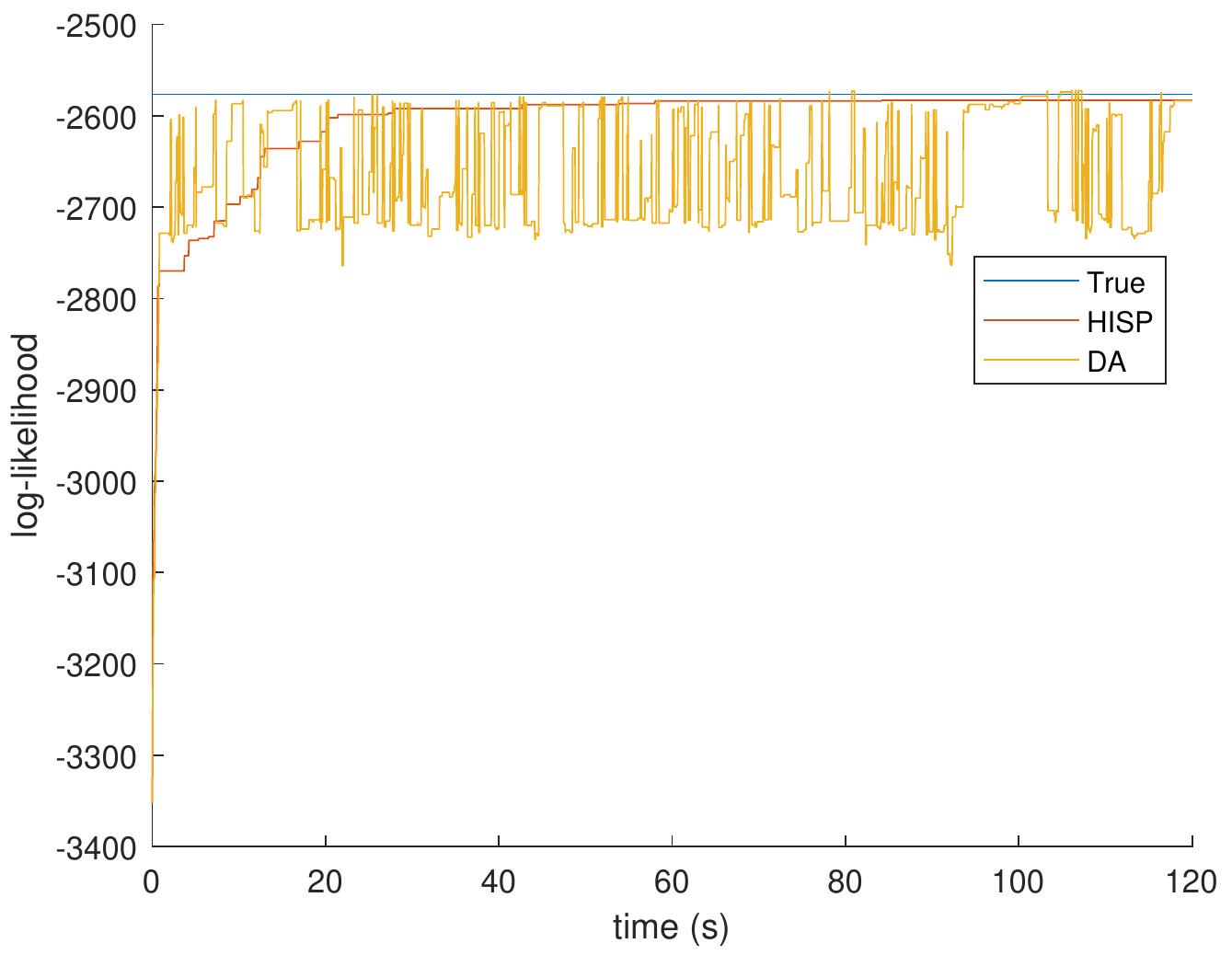}
        \caption{Comparison between different methods.}
        \label{fig:perf_simple}
    \end{subfigure}
    \begin{subfigure}[b]{0.49\textwidth}
        \centering
        \includegraphics[width=\textwidth, trim=95pt 265pt 115pt 280pt, clip]{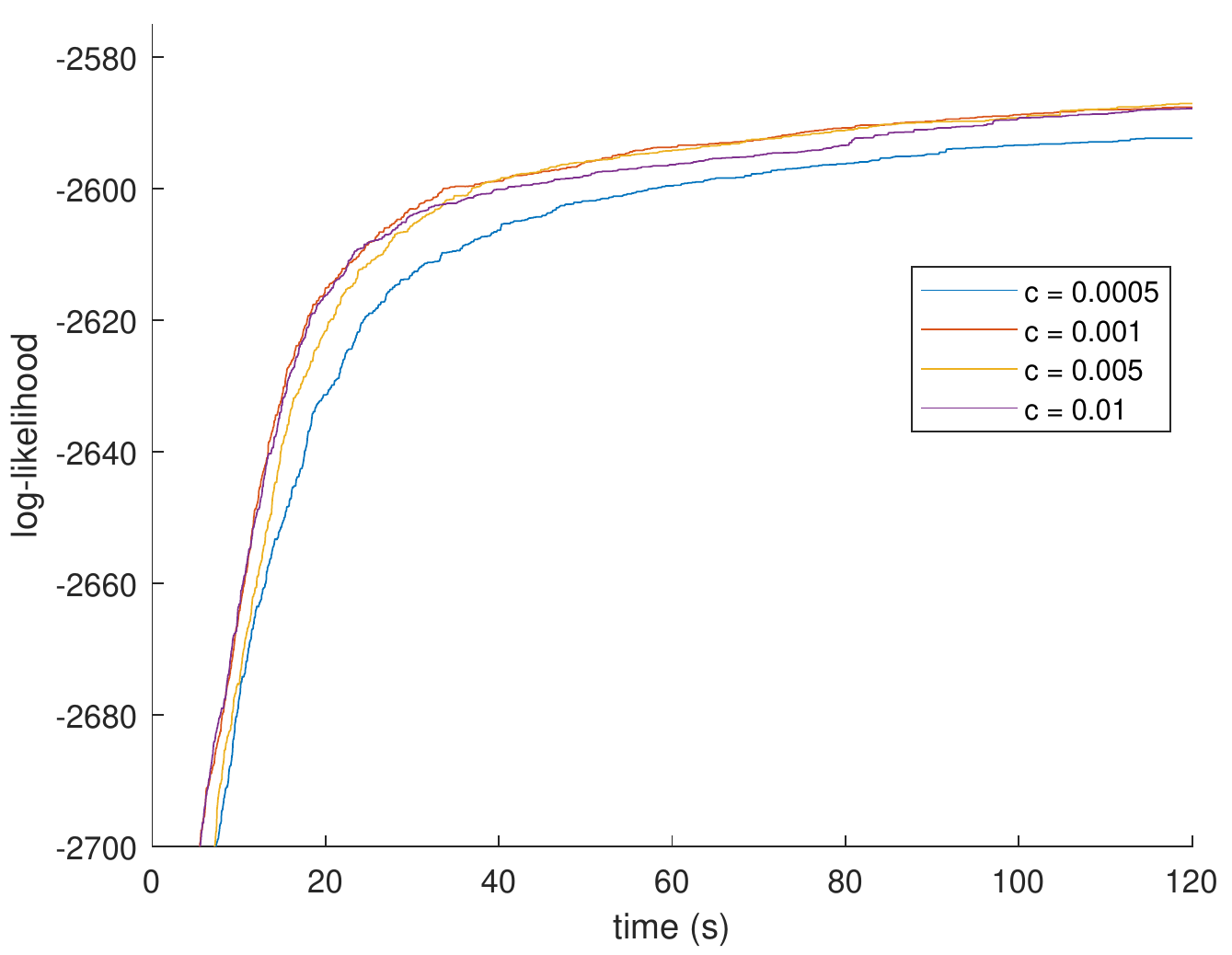}
        \caption{Trace plot for different values of $c$.}
        \label{fig:scen_simple_assess_c}
    \end{subfigure} 
\caption{Simple scenario with performance comparison for different parameter choices.}
\label{fig:scen_simple}
\end{figure}

We first consider a simple scenario, as shown in Figure~\ref{fig:scen_simple_pos}, with $10$ false alarms and $0.1$ appearing objects per time step on average and with a probability of detection of $p_{\mathrm{d}} = 0.9$. The simplicity of the scenario is illustrated in Figure~\ref{fig:scen_simple_dist} where it appears that most of the false alarms are far from any other observation and, conversely, object-originated observations are close to each other. 

The performance of the two considered approaches is first assessed on a single run in Figure~\ref{fig:perf_simple} where the evolution of the log-likelihood is displayed as a function of the computational time. ``HISP'' refers to the proposed approach whereas ``DA'' refers to the MCMC-DA. The difference in behaviour between the proposed approach and MCMC-DA is due to the use of the simulated annealing in the former. Both methods provide satisfactory results in this case and the MCMC-DA's chain mixes well. Figure~\ref{fig:scen_simple_assess_c}, which displays the performance averaged over $50$ repeats, shows that setting the parameter $c$ in the inverse temperature $\rho_t$ to $0.001$ provides the best performance throughout the duration of the runs.

\subsubsection{Scenario with high false-alarm rate}

\begin{figure}
\centering
    \begin{subfigure}[b]{0.49\textwidth}
        \centering
        \includegraphics[width=\textwidth, trim=100pt 265pt 115pt 280pt, clip]{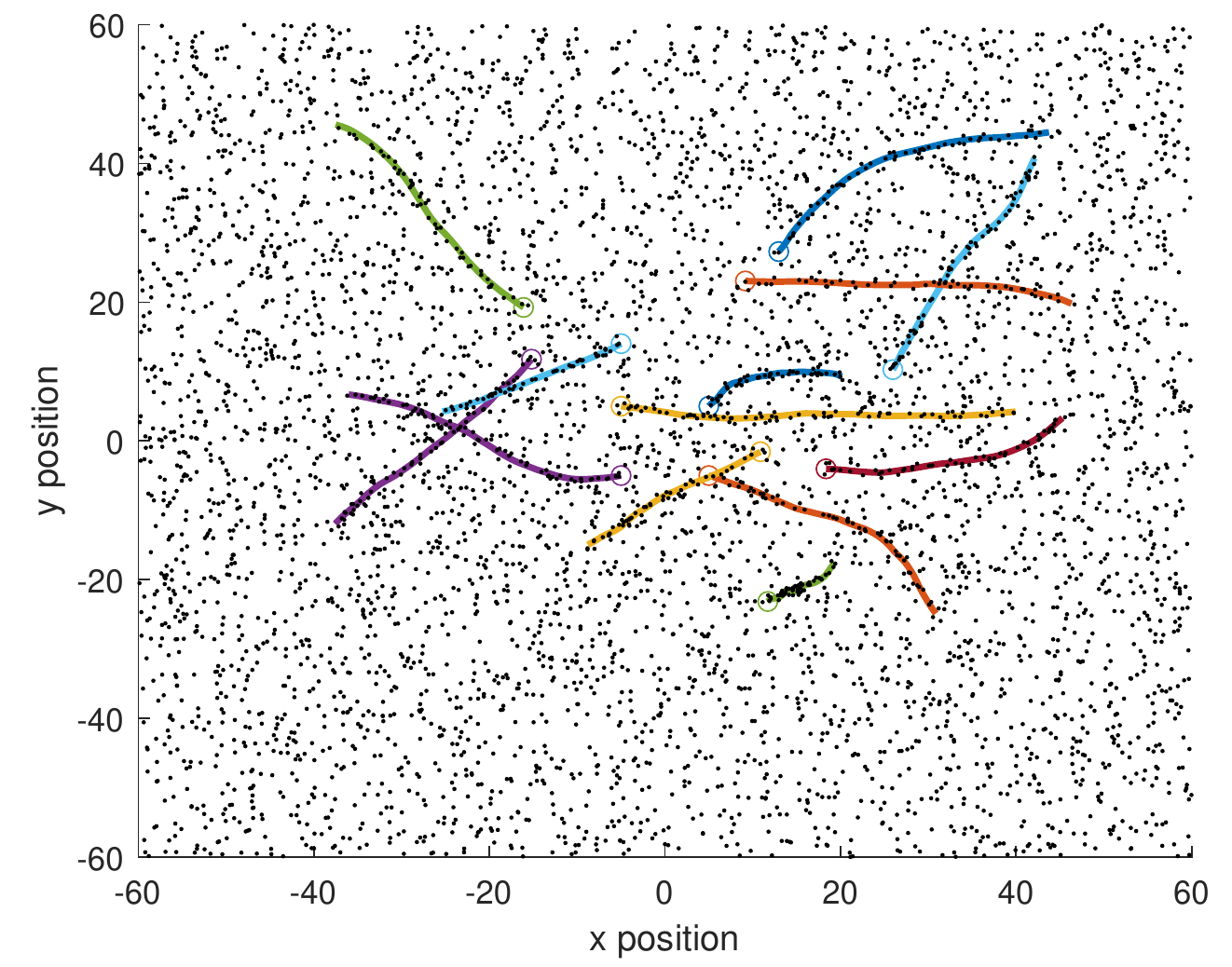}
        \caption{Trajectories (with circles indicating initial position) and observations (black dots).}
        \label{fig:scen_full_pos}
    \end{subfigure}
    \begin{subfigure}[b]{0.49\textwidth}
        \centering
        \includegraphics[width=\textwidth, trim=100pt 265pt 115pt 280pt, clip]{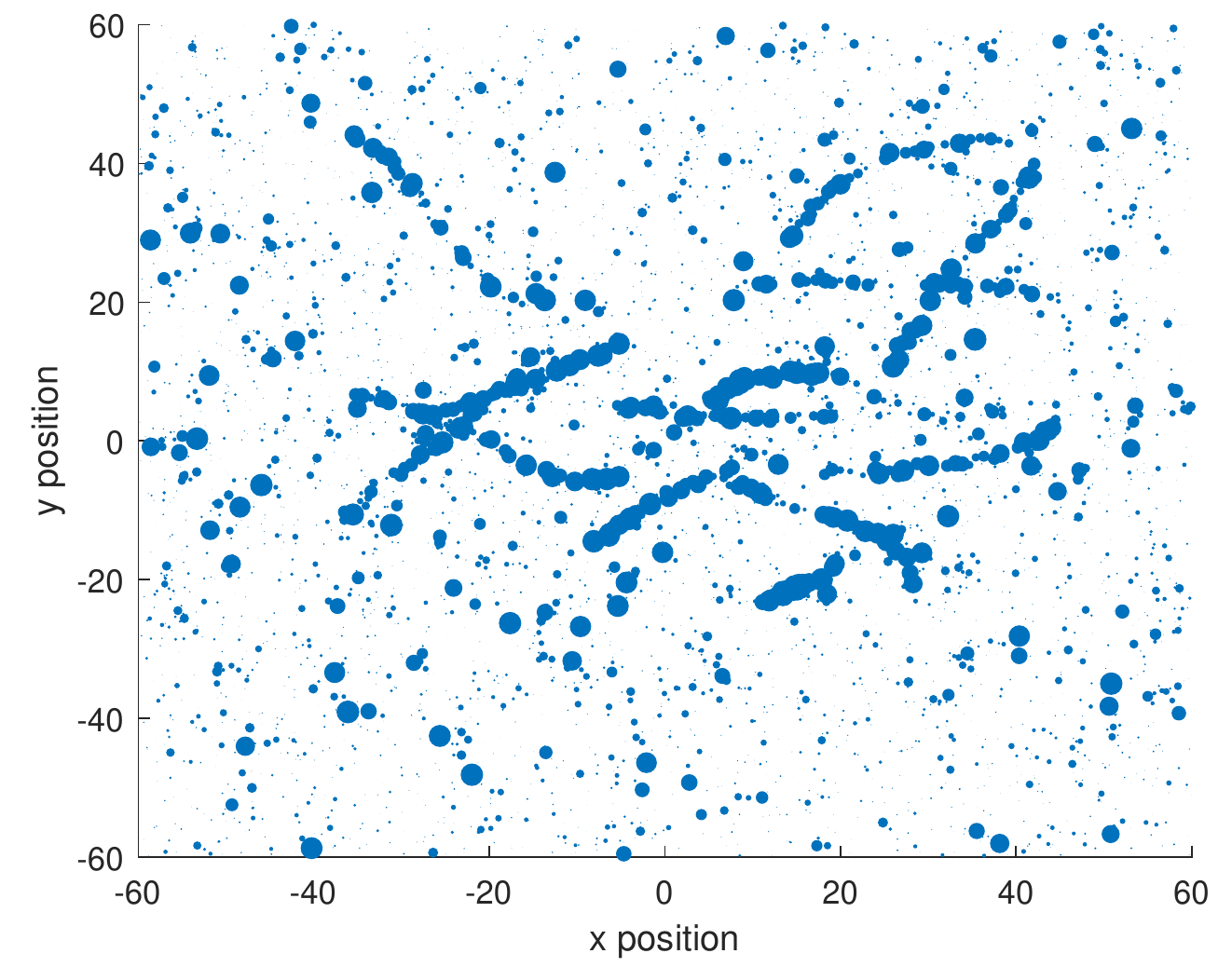}
        \caption{Distance from one given observation to the nearest observation.}
        \label{fig:scen_full_dist}
    \end{subfigure}
    \begin{subfigure}[b]{0.49\textwidth}
        \centering
        \includegraphics[width=\textwidth, trim=100pt 265pt 115pt 275pt, clip]{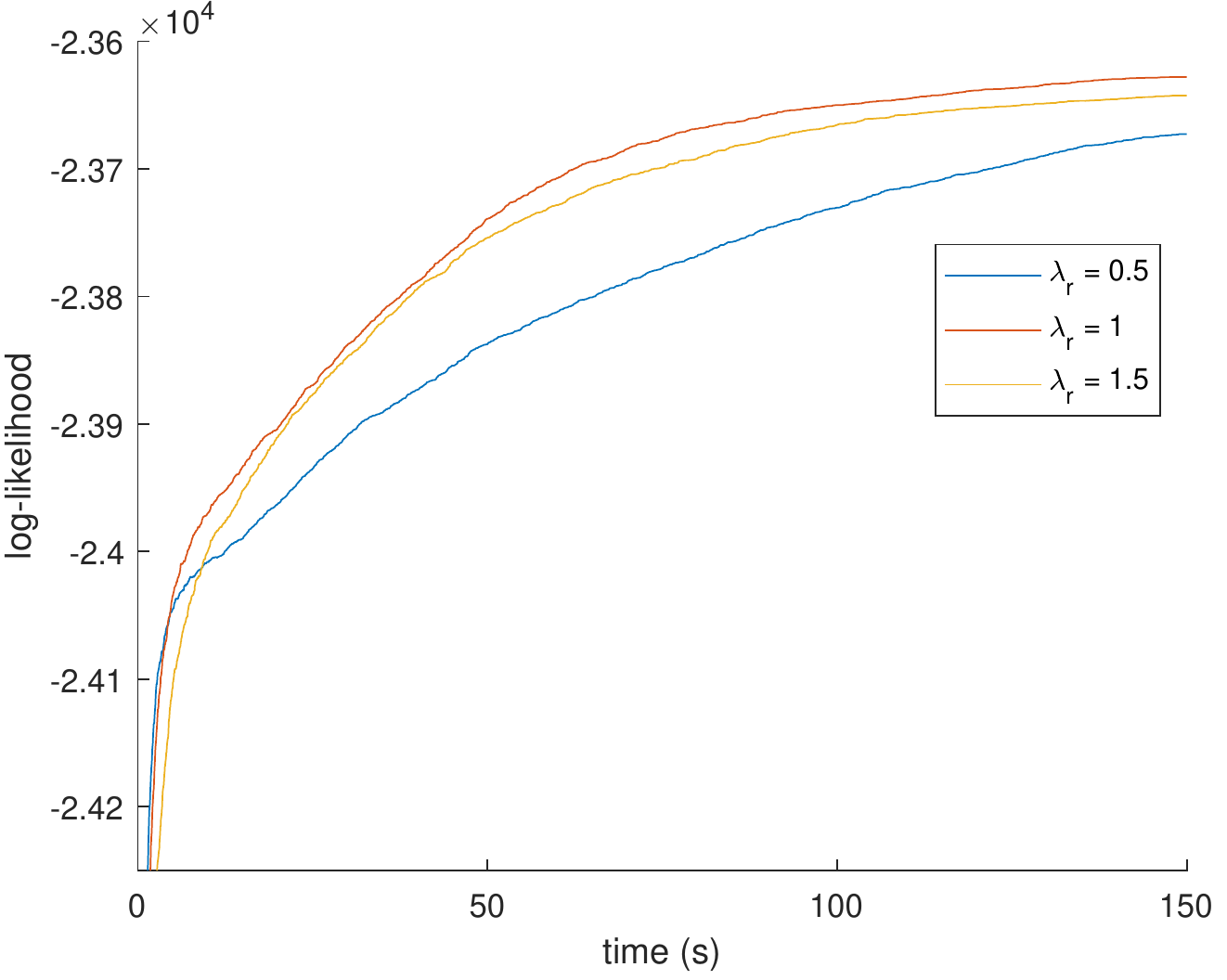}
        \caption{Trace plot for $\lambda_{\mathrm{r}} \in \{0.5,1,1.5\}$.}
        \label{fig:scen_full_assess_lambda}
    \end{subfigure}
    \begin{subfigure}[b]{0.49\textwidth}
        \centering
        \includegraphics[width=\textwidth, trim=100pt 265pt 115pt 275pt, clip]{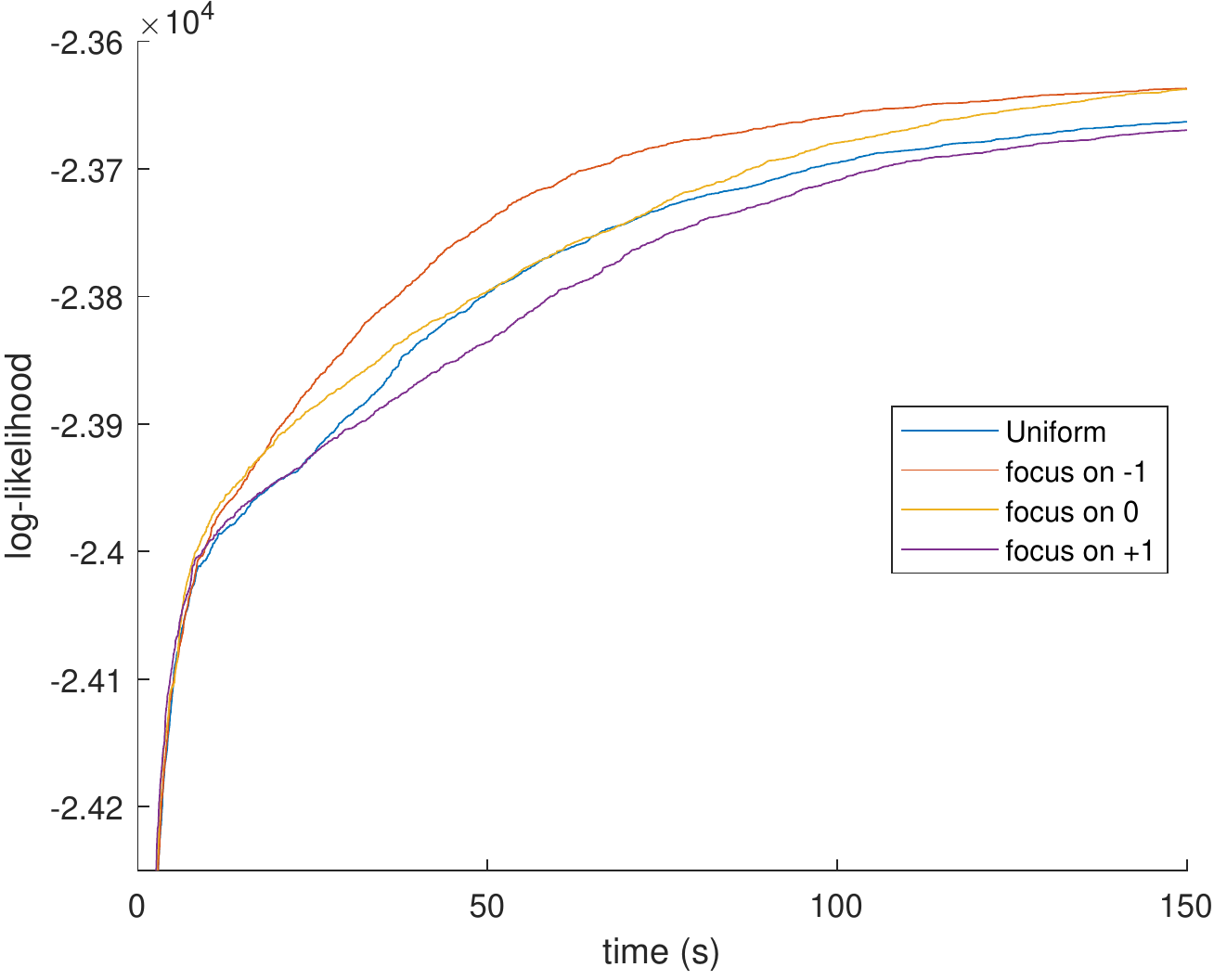}
        \caption{Trace plot for different choices for $\tilde{p}_{\mathrm{c}}$.}
        \label{fig:scen_full_assess_tildePc}
    \end{subfigure}
    \begin{subfigure}[b]{0.49\textwidth}
        \centering
        \includegraphics[width=\textwidth, trim=95pt 265pt 115pt 280pt, clip]{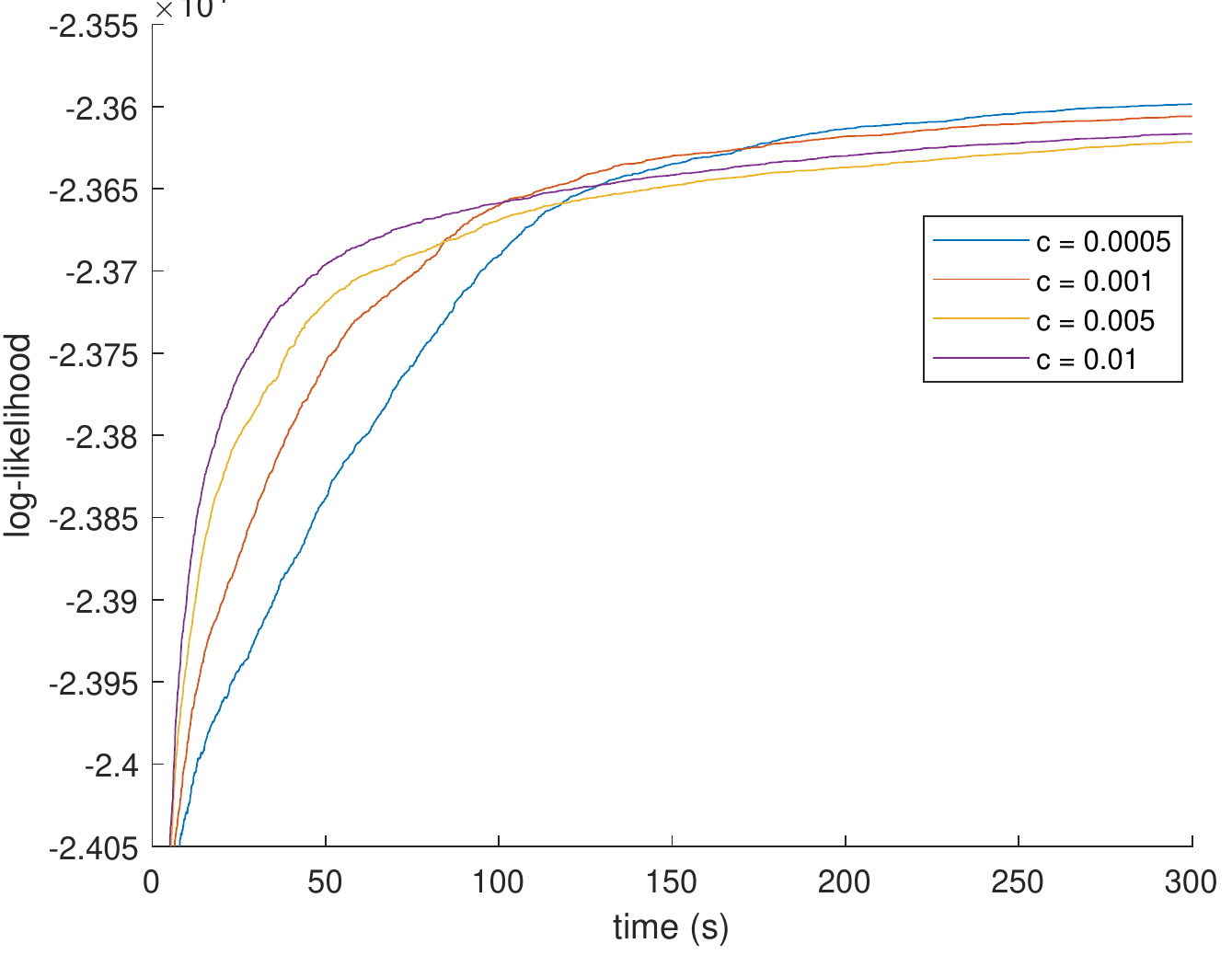}
        \caption{Trace plot for different values of $c$.}
        \label{fig:scen_full_assess_c}
    \end{subfigure} 
    \begin{subfigure}[b]{0.49\textwidth}
        \centering
        \includegraphics[width=\textwidth, trim=100pt 265pt 115pt 275pt, clip]{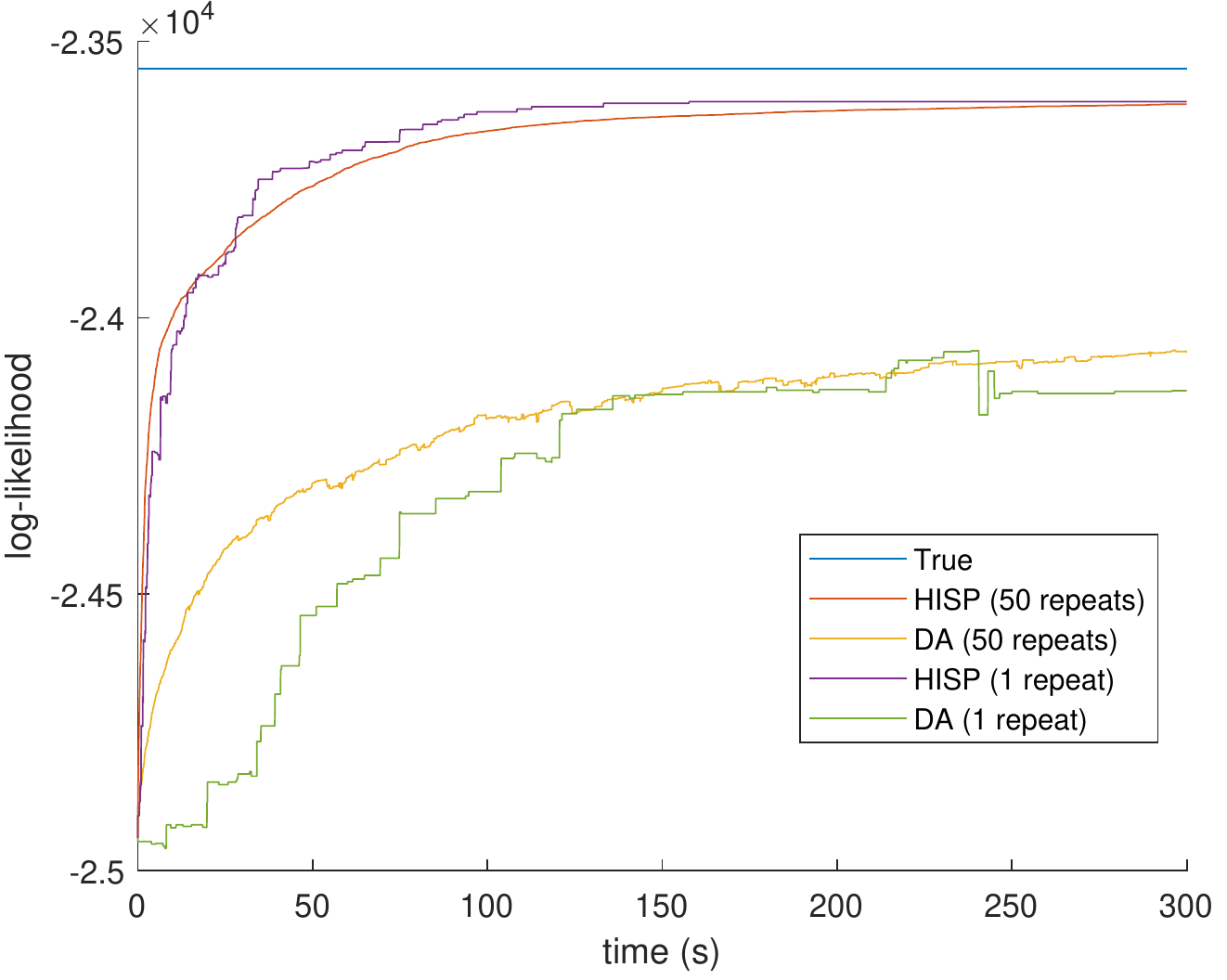}
        \caption{Comparison between different methods.}
        \label{fig:perf_full}
    \end{subfigure}
\caption{Scenario with high false-alarm rate.}
\label{fig:scen_full}
\end{figure}

We consider a first type of challenging scenario, depicted in Figure~\ref{fig:scen_full_pos}, with the following challenging characteristics: there are $100$ false alarms and $0.5$ appearing objects per time step on average and the probability of detection $p_{\mathrm{d}}$ is equal to $0.8$. In this case, it is the large number of false alarms that make the estimation difficult due to the fact that they are likely to form coherent observation sequences over $2$ to $3$ time steps. This aspect is illustrated in Figure~\ref{fig:scen_full_dist} where many false alarms can be seen to be near other observations. Figure~\ref{fig:scen_full_assess_lambda} considers different choices for the Poisson parameter $\lambda_{\mathrm{r}}$ with the log-likelihood being once again averaged over $50$ repeats. The choice $\lambda_{\mathrm{r}} = 1$ allows for rapidly creating tracks while proposing the simultaneous reassignment of 2 tracks often enough to prevent track fragmentation, whereas setting $\lambda_{\mathrm{r}}$ to $0.5$ or $1.5$ does not perform as well. Finally, a few options are compared in Figure~\ref{fig:scen_full_assess_tildePc} for the distribution $\tilde{p}_{\mathrm{c}}$, with the log-likelihood being averaged over $50$ repeats. The assessed options are
$$
\tilde{p}_{\mathrm{c}}( (-1, 0,+1) \given N_{\mathrm{r}}) =
\begin{cases*}
(1/3,\, 1/3,\, 1/3) & as ``uniform'' \\
(1/2,\, 1/4,\, 1/4) & as ``focus on $-1$''\\
(1/4,\, 1/2,\, 1/4) & as ``focus on $0$''\\
(1/4,\, 1/4,\, 1/2) & as ``focus on $+1$'',
\end{cases*}
$$
where $\tilde{p}_{\mathrm{c}}( (\delta_1, \delta_2, \delta_3) \given N_{\mathrm{r}}) = (p_1, p_2, p_3)$ is a shorthand notation for $\tilde{p}_{\mathrm{c}}(\delta_i \given N_{\mathrm{r}}) = p_i$ for $i \in \{1,2,3\}$. The results in Figure~\ref{fig:scen_full_assess_tildePc} show that focusing on $\delta = -1$ yields a slightly better performance, followed by focusing on $\delta = 0$. Once again, this can be attributed to the reduction in track fragmentation. The influence of the parameter $c$ is considered once more in Figure~\ref{fig:scen_full_assess_c} where it appears that $c = 0.0005$ gives the best long-run performance. However, $c = 0.001$ still provides good performance throughout the run time and is considered for the other simulations. Figure~\ref{fig:perf_full} compares the performance of the propose approach with MCMC-DA and shows that the latter does not mix as well as in the first scenario and fails to identify most of the tracks. The fact that the proposed approach does not reach the true log-likelihood can be attributed to local maxima in the posterior possibility function $\Pi$ as well as to identifiability issues. The trace plots are shown for $1$ repeat as well as for $50$ repeats in order to show that the low average performance of the MCMC-DA is not due to averaging.

\subsubsection{Scenario with low probability of detection}

\begin{figure}
\centering
    \begin{subfigure}[b]{0.49\textwidth}
        \centering
        \includegraphics[width=\textwidth, trim=100pt 265pt 115pt 280pt, clip]{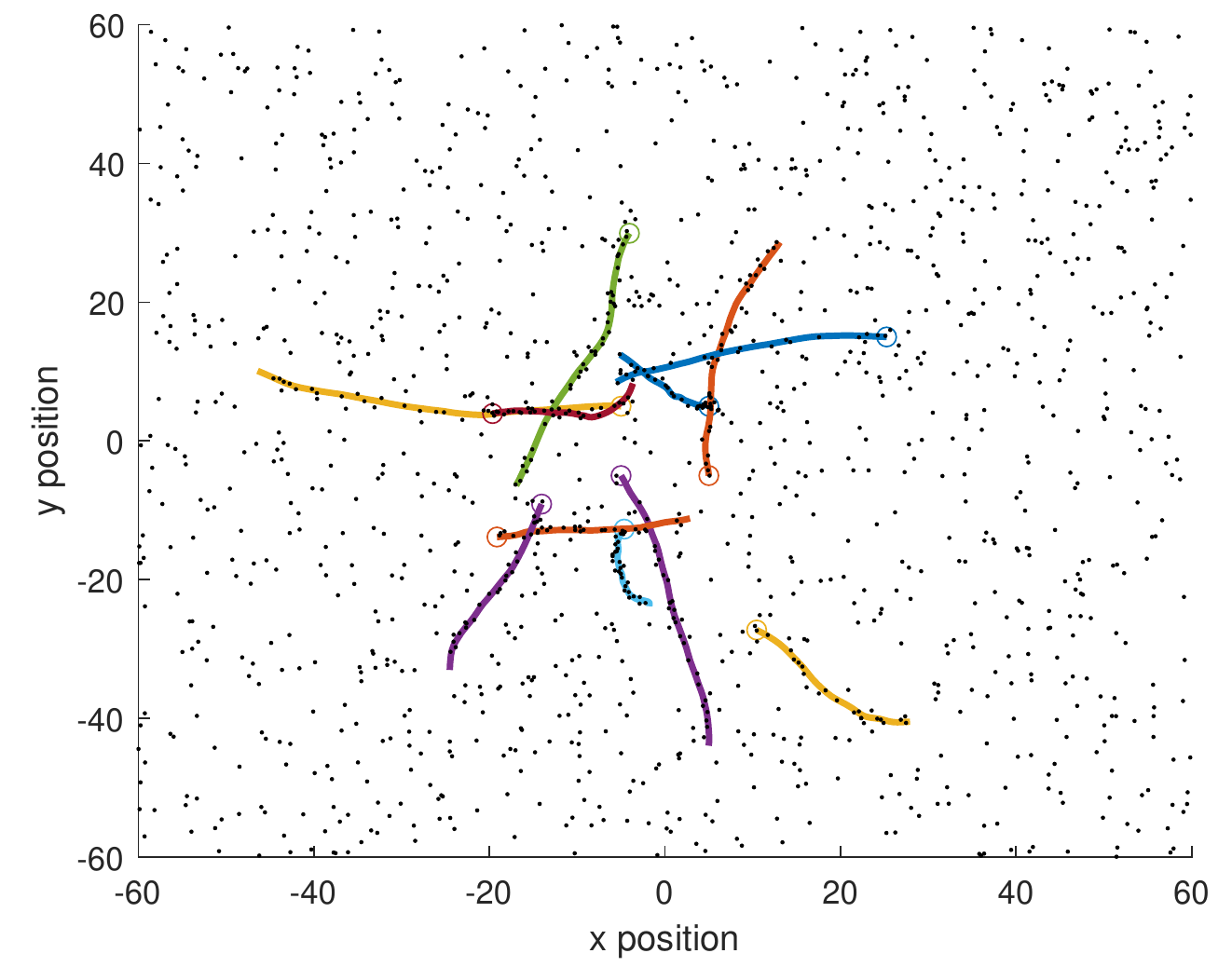}
        \caption{Trajectories (with circles indicating initial position) and observations (black dots).}
        \label{fig:scen_sparse_pos}
    \end{subfigure}
    \begin{subfigure}[b]{0.49\textwidth}
        \centering
        \includegraphics[width=\textwidth, trim=100pt 265pt 115pt 280pt, clip]{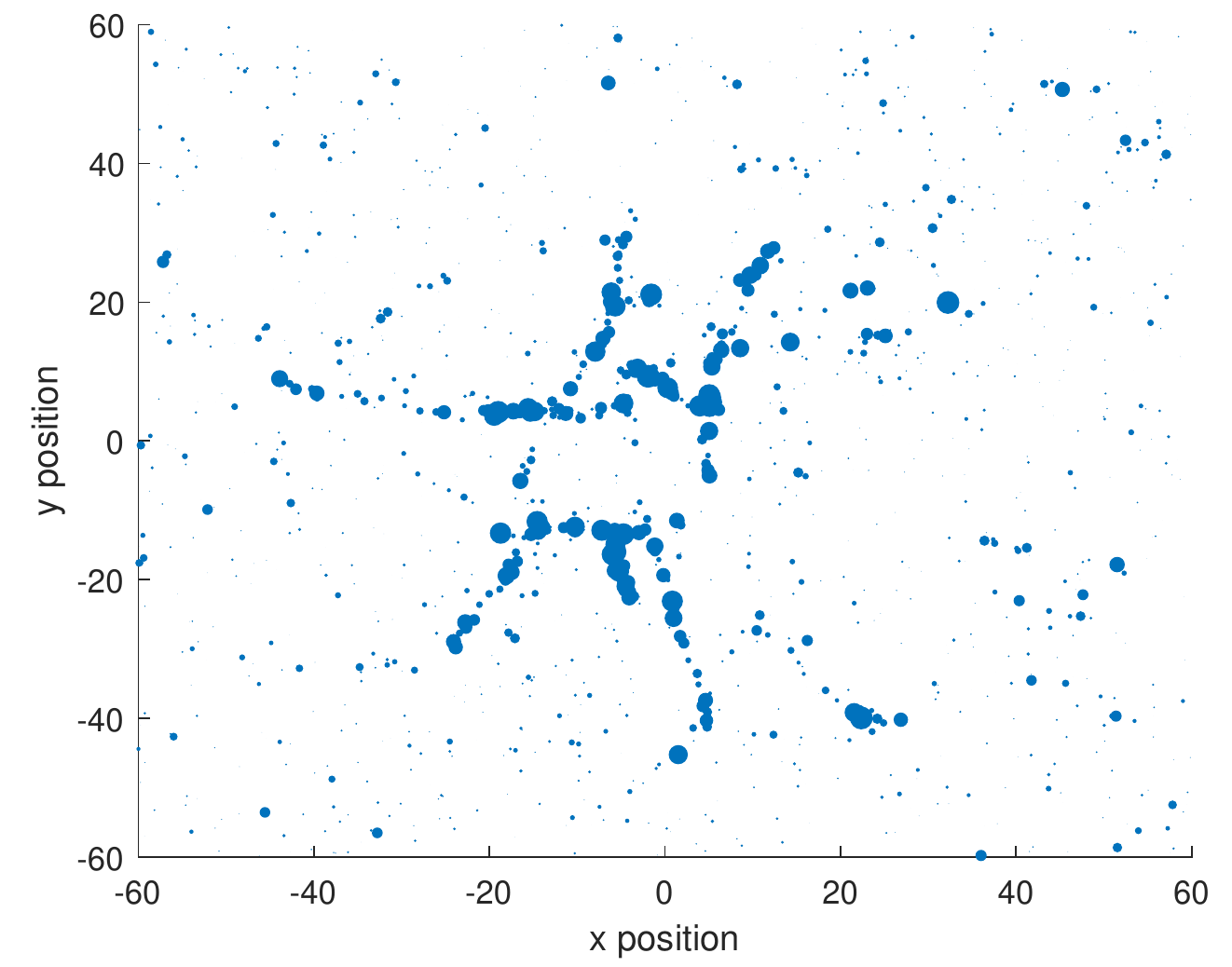}
        \caption{Distance from one given observation to the nearest observation.}
        \label{fig:scen_sparse_dist}
    \end{subfigure}
    \begin{subfigure}[b]{0.49\textwidth}
        \centering
        \includegraphics[width=\textwidth, trim=100pt 265pt 115pt 280pt, clip]{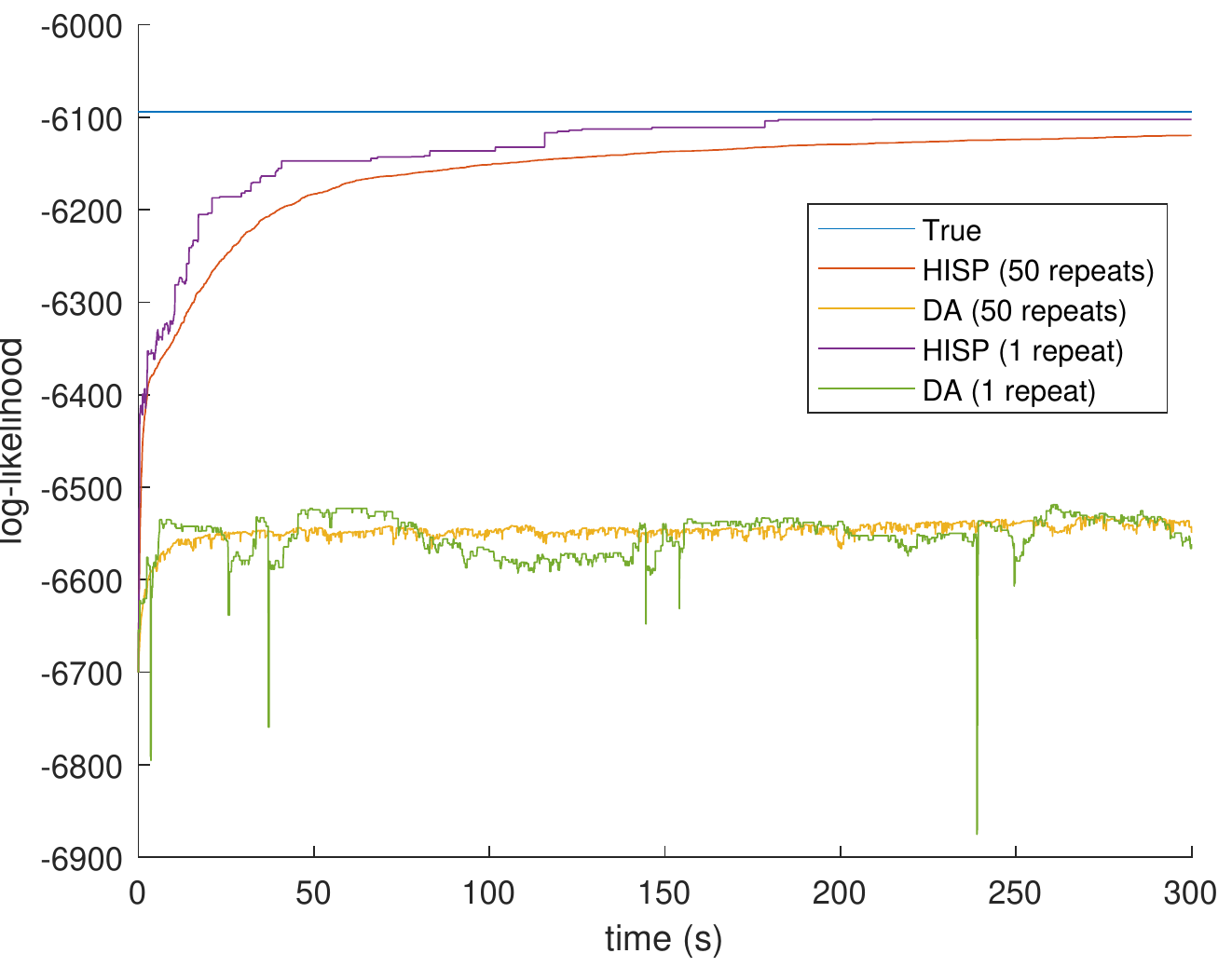}
        \caption{Comparison between different methods.}
        \label{fig:perf_sparse}
    \end{subfigure}
\caption{Scenario with low probability of detection.}
\label{fig:scen_sparse}
\end{figure}

To further assess the performance of the considered approach, we consider another challenging scenario, as shown in Figure~\ref{fig:scen_sparse_pos}, with the following characteristics: there are $25$ false alarms and $0.5$ appearing objects per time step on average and the probability of detection $p_{\mathrm{d}}$ is equal to $0.5$. The difficulty of this scenario is illustrated in Figure~\ref{fig:scen_sparse_dist} where it appears that the inter-observation distance is not sufficient to clearly identify the objects; in particular, the observations belonging to the object at the bottom right barely appear in Figure~\ref{fig:scen_sparse_dist}, emphasising the fact that a probability of detection of $0.5$ is not sufficient to guarantee the spatio-temporal consistency between observations. Figure~\ref{fig:perf_sparse} shows that the proposed approach can capture most of the structure of the scenario whereas the MCMC-DA did not identify the majority of tracks in the allocated time. 

\bibliographystyle{abbrv}
\bibliography{main}

\end{document}